\documentclass[a4paper,twocolumn,11pt,accepted=2023-11-24]{quantumarticle}
\pdfoutput=1

\usepackage[sort&compress, numbers]{natbib}
\usepackage{amssymb}
\usepackage{amsmath}
\usepackage{graphicx}
\usepackage{amsfonts}
\usepackage[usenames, dvipsnames]{xcolor}
\usepackage{bm}
\usepackage{bbm}
\usepackage{color}
\usepackage{hyperref}
\hypersetup{pdfnewwindow=true,
            colorlinks=true,
            linkcolor=blue,
            citecolor=blue,
            urlcolor=blue}
\usepackage{comment}
\usepackage{soul}
\usepackage[normalem]{ulem}
\usepackage[caption=false, justification=centerlast]{subfig}

\usepackage[utf8]{inputenc}
\usepackage{amssymb}
\usepackage{amsmath,graphicx, dsfont, bm}
\usepackage[normalem]{ulem}
\usepackage{diagbox}
\usepackage[dvipsnames]{xcolor}
\usepackage{ifsym}
\usepackage{physics}

\newcommand{\mean}[1]{\langle {#1} \rangle}
\newcommand{\prjct}[1]{\mathinner{|{#1}\rangle}\!\!\mathinner{\langle{#1}|}}

\newcommand{\id}{\mathds{1}}
\newcommand{\ii}{\mathrm{i}}

\newcommand{\cE}{\mathcal{E}}
\newcommand{\cF}{\mathcal{F}}

\newcommand{\cL}{\mathcal{L}}
\newcommand{\cI}{\mathcal{I}}

\renewcommand{\t}[1]{\mathrm{#1}}

\newcommand{\be}{\begin{equation}}
\newcommand{\ee}{\end{equation}}

\newcommand{\bs}{\begin{split}}
\newcommand{\es}{\end{split}}

\newcommand{\BS}{\begin{equation}\begin{split}}
\newcommand{\ES}{\end{split}\end{equation}}

\begin{document}
\title{Optimal nonequilibrium thermometry 
in Markovian environments}

\author{Pavel Sekatski}
\affiliation{Department of Applied Physics, University of Geneva, 1211 Geneva, Switzerland}
\author{Mart\' i Perarnau-Llobet}
\affiliation{Department of Applied Physics, University of Geneva, 1211 Geneva, Switzerland}

\begin{abstract}
    What is the minimum time required to take a temperature? In this paper, we solve this question for  a large class of processes  where temperature is inferred by measuring a probe (the thermometer)  weakly coupled to the sample of interest, so that the probe's evolution is well described by a quantum Markovian master equation. Considering the most general control strategy on the probe  (adaptive measurements, arbitrary control on the probe's state and Hamiltonian), we provide  bounds on the achievable measurement precision  in a finite amount of time,  and show that in many scenarios these fundamental limits can be saturated with a relatively simple experiment.  
    We find that for a general class of sample-probe interactions the scaling of the measurement uncertainty is inversely proportional to the time of the process, a shot-noise like behaviour that arises due to the  dissipative nature of thermometry.
   As a side result, we show that the Lamb shift induced by the probe-sample interaction can play a relevant role in thermometry, allowing for finite measurement resolution in the low-temperature regime. More precisely, the measurement uncertainty decays polynomially with  the temperature as $T\rightarrow 0$, in contrast to the usual exponential decay with $T^{-1}$. 
    We illustrate these general results for (i) a qubit probe interacting with a bosonic sample, where the role of the Lamb shift is highlighted, and (ii) a collective superradiant coupling between a $N$-qubit probe and a sample, which enables a quadratic  decay with $N$ of the measurement uncertainty.
\end{abstract}

\maketitle

\section{Introduction}




Temperature is a physical quantity important in most natural sciences and some aspects of daily life. The task of thermometry -- measuring the temperature -- is essential in physics, chemistry, medicine, and cooking, to name a few. 
A good thermometer should be accurate (on average it provides the right temperature) and precise (fluctuations around the true value are small). The former requires an accurate design of the thermometer, whereas the latter demands 
collecting enough statistics: either by using a macroscopic thermometer where each subsystem equilibrates with the sample, or by repeating the experiment enough times when dealing with thermometers at the  quantum scale~\cite{Giazotto_2006,Yue_2012, Mehboudi_2019rev,Carlos_2016,DePasquale2018}. Such processes take time, and in fact we also wish a good thermometer to be fast. Motivated by this practical limitation,  in this paper we derive   fundamental limits on  finite-time thermometry, and discover that in some situations these limits can be saturated with a relatively simple experiment. 
 

Concretely, we consider a probe (acting as a thermometer)
in contact with a sample -- a reservoir at temperature~$T$. Their interaction leaks information on the sample's temperature (or another physical property) to the state of the probe. After some time, the latter is measured in order to estimate the parameter $T$. To benchmark the quality of such estimation schemes we consider the quantum Fisher information (QFI) of the final state of the probe $\mathcal{F}$~\cite{Braunstein_1994}.

The most natural starting point is equilibrium thermometry, where the probe is given sufficient time to thermalize to its Gibbs state, $\rho = e^{-\beta H}/\tr e^{-\beta H}$ with $\beta=1/k_{\rm B} T$ and $k_{\rm B}$ the  Boltzmann constant ($k_{\rm B}=1$ in what follows)~\cite{Paris_2015,Correa_2015,Guo_2015,Campbell2017,Campbell2018,Plodzie2018}.
Remarkably, the (quantum) Fisher information of the Gibbs state is proportional to the heat capacity of the system, because the average energy turns out to be the optimal estimator for temperature as it saturates the Cramer-Rao bound~\cite{Ruppeiner1979,Old_PRE,Paris_2015,Correa_2015}. This  provides a practical recipe to design good equilibrium thermometers: the probe's Hamiltonian $H$ needs to  be engineered to maximise the heat capacity of the probe around the estimated temperature $T$. For example, this can be achieved by 
preparing the probe close to a critical point, see some examples in \cite{Mehboudi_2019rev,SaladoMeja2021}. 
The fundamental limits of equilibrium thermometry where derived by Correa \emph{et al} in~\cite{Correa_2015} by maximising the QFI with respect to probe's Hamiltonian, 
showing in particular that it can scale 
at most quadratically with the number of subsystems in the probe~\cite{Reeb2014,Plodzie2018}. 

Nonequilibrium thermometry offers an alternative framework to enhance the QFI of the probe. In this case, the probe is measured before reaching equilibrium, so that extra information can be potentially achieved. This requires  knowledge of the dissipative dynamics and precise timing for the measurement, but can lead to substantially higher QFI than standard equilibrium thermometry \cite{Brunelli_2011, Brunelli_2012, Jevtic_2015,Guo_2015,Tham2016,Pasquale_2017,Hofer2017b, Cavina_2018,RomnAncheyta2019,Mitchison_2020, Mancino_2020,Jrgensen2020,Seah2019,Shu2020,Henao_2020,Bouton_2020,Hovhannisyan2021,mitchison2021taking}.  
The maximum achievable QFI in any such scenario  has been recently derived in~\cite{Hovhannisyan2021} (note that, in contrast to the present work, the sensing time is not taken to be a resource in~\cite{Hovhannisyan2021}).   

These  results are promising for the design of quantum thermometers. However, in most cases they disregard the importance of the sensing time, i.e. how long it takes to reach the desired state. A clear example is equilibrium thermometry: whereas critical systems have a higher heat capacity (i.e. QFI), they also  require longer thermalisation times \cite{huang2009introduction}. There is hence an interesting tradeoff between the probe's QFI and the required time of the estimation process. In fact, in analogy with frequency estimation \cite{Bollinger1996,Huelga1997}, it is natural to consider both the number of probes and the sensing time are resources for thermometry.


To take time into account, in this article we consider the QFI rate as the figure of merit -- the QFI of the final state divided by the total interaction time~\cite{Correa_2015,Mehboudi_2019rev}.  
Considering the most general control strategy on the probe (adaptive measurements, control on the probe's state and Hamiltonian), we derive upper bounds on the QFI rate whenever the thermalisation process is well described by a  general class of Markovian master equations  
by exploiting  techniques derived in \cite{ffqc,Rafal2017,zhou2018} in the context of frequency estimation. Using Markovian master equations for non-equilibrium thermometry is ubiquiquituous in theoretical works (see e.g. the review \cite{Mehboudi_2019rev}), but also of crucial experimental relevance: recent experiments in non-equilibrium thermometry involving single-atom probes for quantum gases~\cite{Bouton_2020} can be well described by this framework~\cite{nettersheim2022sensitivity}.  

Our main results and their implications can be summarised as follows:
\begin{itemize}
    \item We show that, for a large class of Markovian dynamics, 
    the QFI grows  at most linearly with the total sensing time $\tau$, a shot-noise like behaviour common of dissipative processes. More precisely, the QFI  $\mathcal{F}$ is upper bounded by an inequality of the form $\mathcal{F} \leq f(\mathcal{L}) \tau$, see Eq. \eqref{eq: QFI bound final main} and \eqref{eq: HLS decomposition}, where  $f(\mathcal{L})$ is a function that depends on the specific Linblad evolution. The bounds  can be easily computed from the jump operators and the corresponding rates without the need to solve the dynamics.
     \item We  apply these bounds to Lindblad master equations  derived from  a probe-sample weak interaction of the form $H_I=A\otimes B$, with $A$ ($B$) acting on the probe (sample). Then, we show that for arbitraty probe Hamiltonian $H$ our bounds can be put in a very intuitive form (see Eq.~\eqref{eq:upboundO}):
   \begin{align}
      \mathcal{F} \leq  \tau \tilde{f}(B) (\Delta A)^2
    \end{align}
    where $\Delta(A) = \lambda_\t{max} - \lambda_\t{min}$ is the spectral gap of the operator $A$ (the difference between its  maximal and minimal eigenvalues), and $\tilde{f}(B)$ is a function that depends on the spectral density of the sample. This can be understood as a speed limit relating the amount of information that the probe can acquire about the sample with their coupling given by~$H_I=A\otimes B$. We also show how to engineer the probe Hamiltonian $H$ in order to attain the bound.
      \item 
    In absence of temperature dependent Lamb shift, these upper bounds on the QFI can  be saturated by a fast measure-and-prepare strategy. The optimality of these strategies were already noted numerically for qubit probes \cite{Correa_2015,Mehboudi_2019rev}, and our results generalise these considerations to arbitrary Markovian evolutions. We also develop an autonomous scheme where the fast measure-and-prepare  strategy is achieved through coupling the probe to an external zero-temperature bath, and monitoring the energy flux from the sample to the bath. 
    \item The fast measure-and-prepare strategy can in principle be improved in the presence of a temperature-dependent Lamb shift.   
    Whereas this contribution is usually negligible compared to that obtained from the temperature-dependent  decay rates, 
    we argue that it plays a dominant role at low temperatures. By focusing on a qubit probe with frequency $w$ interacting with a bosonic sample,
    we show that the QFI can scale as  $\mathcal{F} \propto T^{1+\alpha}$ as $T\rightarrow 0$ ($\alpha$ is the Ohmicity of the sample), in contrast to the standard exponential $\mathcal{F} \propto e^{-w/T}$ obtained when the Lamb shift is neglected~\cite{Paris_2015,Correa_2017}. Similar polynomial decays had been obtained in the steady state of strongly coupled sample-probes states \cite{Correa_2017,Hovhannisyan_2018,Potts_2019,Jrgensen2020}.  
    To  exploit this advantage, we show that the probe has to be left to evolve coherently for a finite time analogously to frequency estimation scenarios (see in this context \cite{Mitchison_2020,mitchison2021taking}),  which is markedly different from the  fast  measure-and-prepare strategy.
    
    \item To illustrate the general applicability of our framework, we also develop optimal thermometric protocols for more complex probes, consisting of a $N$ qubits  coupled to the sample through a collective coupling \cite{dicke54,Latune2020}. In this case, we show that the QFI can scale as $N^2$, and characterise the experimental requirements to obtain this enhanced scaling.   
\end{itemize}
It is also important to emphasise that the bounds and measurement schemes we  derive are not limited to the task of thermometry. They apply to any scenario where the probe's interaction with the sample is well described by a Markovian master equation for the state of the probe, with the jump rates and the Lamb shift depending on some physical property of the sample. 
Our bounds only rely on the form of this master equation, and hence set a general (and in many cases attainable) limit on how well a physical property of the sample can be estimated through the probe in a finite time. As an illustration we consider the estimation of the Ohmicity of the bath in sec.~\ref{sec: Ohmicity}.

The paper is structured as follows. In Sec. \ref{sec: Framework}, we introduce the different techniques and quantities of interest.  The main results, consisting of the differents bounds on QFI as well as optimal measurement strategies, are then presented in Sec. \ref{sec: results}. In Sec. \ref{Sec: Case Studies}, we apply such general considerations to a qubit interacting with a bosonic sample, 
and to a collective coupling between the sample and a $N$-qubit probe. We close in Sec. \ref{sec: outlook and conclusions} with a short summary and a  discussion of future  endeavours.

\begin{figure}
    \centering
    \includegraphics[width=0.9 \columnwidth]{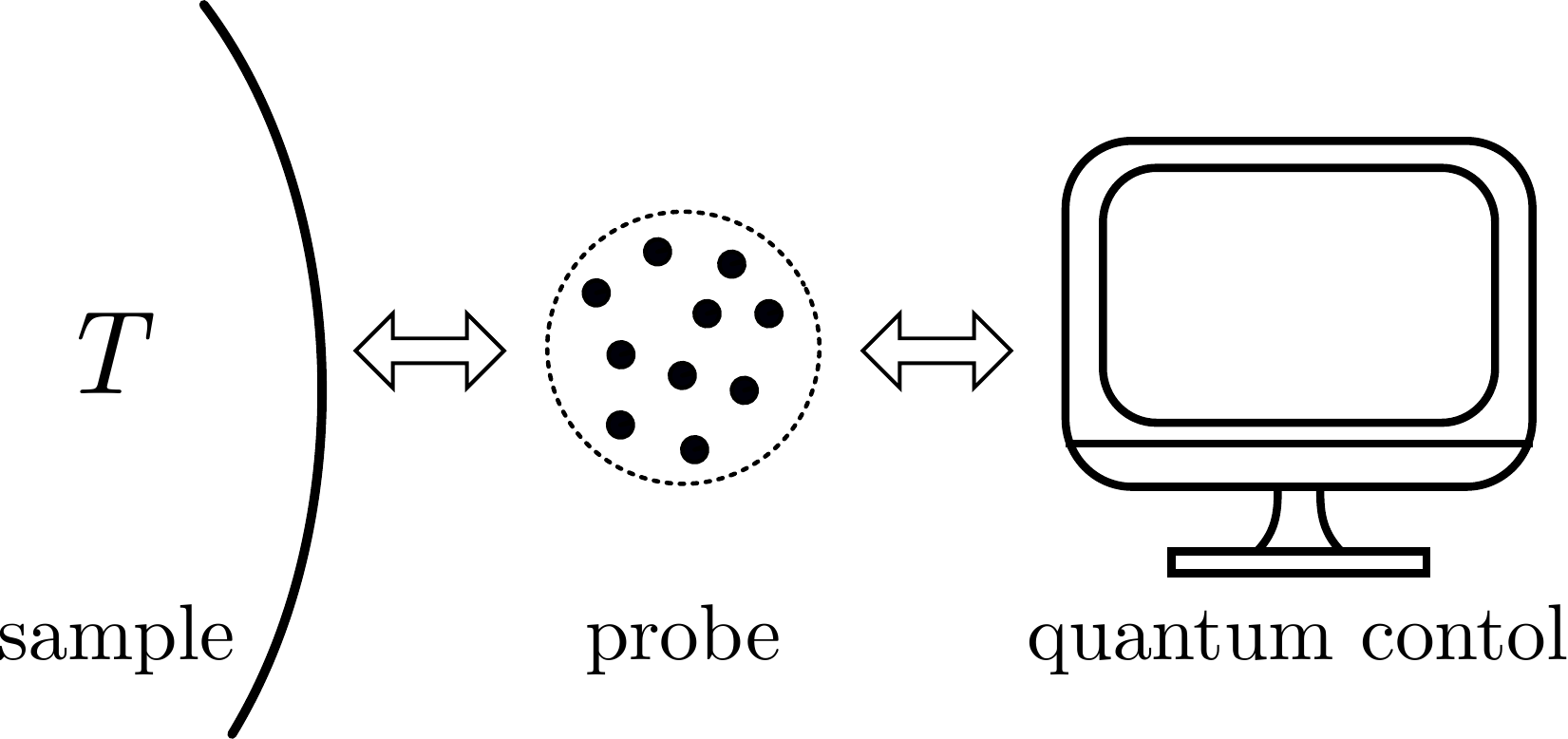}
    \caption{The general sensing scheme we consider. The temperature $T$ (or another parameter) of a large sample at thermal equilibrium is estimated
    by coupling it to a probe 
    for a finite time $\tau$.
    We assume that the probe's evolution 
    is well modeled by Markovian semi-group dynamics.
    In addition, we allow the possibility that probe's dynamics is assisted with general fast quantum control. In particular, this may include continuous measurements, adaptive schemes with feedback,  and entangling gates with auxiliary quantum systems. 
    }
    \label{fig:general setup}
\end{figure}


\section{Framework} 
\label{sec: Framework}

In this section, we 
start by reviewing the basics of parameter estimation, with emphasis on thermometry. We then introduce Lindblad master equations, which are used to describe the thermalisation of the probe when put in contact with the sample, as well as  the  general framework to describe arbitrary quantum control on the probe (including measurements and adaptive strategies), see Fig. \ref{fig:general setup}. Finally, we describe the results of~\cite{ffqc,Rafal2017,zhou2018}, which are the main technical tools used to find optimal finite-time protocols for thermometry. 

\subsection{Parameter estimation and Quantum Fisher Information}

Consider a random variable $X$ collecting the results of all measurements performed during an experiment. To simplify the following discussion we assume that~$X$ is discrete. For thermometry (or any parameter estimation task) the description of the experiment is fully captured by the dependence of~$X$ on the parameter~$T$, as given by the family of probability distributions $p(x|T)$ modelling $X$.  The Fisher information $\mathcal{I}(T)=\mathds{E}\left[ \left(\frac{\partial }{\partial T}\log p(X|T)\right)^2 \Big|T \right] $, associated to $X$ at a given $T$, quantifies how much information  the experiment reveal about the parameter (note that the expected value $H(X) =\mathds{E}\left[ \log p(X|T) \big|T \right]$ is the entropy of $X$).  Denoting $p_x = p(x|T)$ the Fisher information can be easily expressed as
\be\label{eq: Fisher info}
\mathcal{I}(T) = \sum_x \frac{(\dot p_x)^2}{p_x}
\ee
under some regularity conditions\footnote{The expression is not well defined if some of the probabilities $p_x$ are strictly zero.}. Here and in the rest of the paper we use the dot to denote the derivative with respect to the estimated parameter $\dot p_x = \frac{\partial}{\partial T} p_x$. One verifies from Eq.~\eqref{eq: Fisher info} that $\mathcal{I}(T)$ can also be understood geometrically as the temperature susceptibility of the probability measure $\bm p(T) =(p_1,p_2,\dots)$ with respect to Kullback-Leibler divergence (also Hellinger or Bhattacharyya distances). 

Fisher information plays a central role in statistics. In particular, via the famous Cram\'er-Rao bound it sets a limit on how well the parameter can be estimated with an unbiased estimator $\hat T(X)$
\be
\mathds{E}\left[(\hat T-T)^2 \right] \geq \frac{1}{\cI(T)}.
\ee
This bound can also be adjusted to the context of Bayesian inference \cite{gill1995}, where the prior knowledge of the parameter $T$ is explicitly taken into account. Futhermore, for a fixed prior Bernstein–von Mises theorem suggests (under some regularity conditions) that  if the experiment is long enough and done well the posterior distribution is fully determined by $\cI(T)$.

In a quantum experiment the final state of the system  (that can include classical registers) is not described by a probability measure $\bm p(T)$ but by a density matrix~$\rho(T)$. Fixing the measurement performed on the final state relates $\rho(T)$ to a random variable $X$ and its Fisher information $\cI(T)$. It it then natural to define the quantum Fisher information $\cF(T)$ (QFI) of the state $\rho(T)$ as the maximal Fisher information for all possible measurements. With this definition it is clear that the role that $\cI(T)$ plays for parameter estimation is endorsed by QFI in quantum experiments. Remarkably, $\cF(T)$ can be written in a closed form
\be
    \cF_{\rho(T)}= \tr L \dot \rho\\
\ee
with the symmetric logarithmic derivative operator $L$ given by the equation $\frac{1}{2}\{L,\rho\} =\dot \rho$~\cite{Braunstein_1994,Mandelbrot_1956, Uffink_1999}. Analogously to the classical case the QFI can be understood geometrically as temperature susceptibility of the state $\rho(T)$ with respect to Bures distance (or Uhlmann's fidelity). 

The final state of the system depends on how it was prepared at the beginning of the experiment. Just like the final measurement the preparation procedure is in principle under control. It is thus more natural to describe the experiment by a quantum channel
$\cE_T$ mapping the initial state of the system $\varrho_0$ to its final state $\rho(T)= \cE_T(\varrho_0)$, and define the QFI of the channel by optimizing over the preparation
\be
\cF_{\cE_T} = \max_{\varrho_0} \cF_{\rho(T)}.
\ee
As for the states the QFI of a channel can be expressed in a simple form, if one allows trivial extensions of the channel $\cE_T\otimes\t{id}$~\cite{fujiwara2008}. 

Finally, in practice it is important to consider the duration of the experiment $\tau$ as a resource for parameter estimation. To benchmark experiments we analyse the QFI as a function of time $\cF_{\cE_{|T\tau}}$. Intuitively, the longer the experiment the better can one estimate the parameter. For example, because the (quantum) Fisher information is additive for independent random variables  $\cI_{(X,Y)}(T) = \cI_{X}(T)+ \cI_{Y}(T)$ and product states $\cF_{\rho(T)\otimes \sigma(T)} =\cF_{\rho(T)} +\cF_{\sigma(T)}$, it is doubled if an experiment is repeated twice. This observation gives rise to the so called diffusive (or standard quantum limit) scaling~$\cF_{\cE_{\tau|T}}\propto \tau$. Among others it describes experiments where the same preparation and measurement is repeated many times. The best scaling, called ballistic (or Heisenberg), possible in quantum experiments is $\cF_{\cE_{\tau|T}}\propto \tau^2$~\cite{giovannetti2006quantum}. This is for example the case when the dynamics of the system is unitary $\cE_{\tau|T}[\rho_0] = e^{-\ii\,  \tau T\,  H} \rho_0\,  e^{\ii\,  \tau T\, H}$. This gives rise to a faster scaling of precision with which the parameter is estimated  with respect to the sensing time.




\subsection{Probe-Sample interaction as a Lindblad master equation}

In the context of this paper the sample is a large reservoir at thermal equilibrium. Ideally, the state of the reservoir is unaffected by the interaction with a small probe, so that the reservoir can be virtually replaced with a fresh one all the time. In consequence the probe-sample interaction can be naturally described by a time-independent Markovian semi-group evolution of the probe system alone. Thus we assume the dynamics of the probe to be given by a master equation of the form 
\be\label{eq: diss}\begin{split}
    \frac{\dd}{\dd t} \rho =\mathcal{L}(\rho) &=  - \ii [H + H_{{LS}},\rho ] + \mathcal{D}(\rho), \\
    \t{with} \qquad
    \mathcal{D}(\rho) &= \sum_\omega \gamma_\omega \left( A_\omega \rho A_\omega^\dag- \frac{1}{2}\{ A_\omega^\dag A_\omega,\rho\} \right).
\end{split}
\ee
We assume, that the dissipator $\mathcal{D}(\rho)$ is of the Lindblad form with respect to the jump operators $A_\omega$ that do not depend on bath temperatures (but the rates $\gamma_\omega$ do). Furthermore, the jump operators satisfy
\be\label{eq: jump operators 2}
A_\omega = \sum_{(\varepsilon',\varepsilon) \in P_\omega}  \Pi_\varepsilon A_\omega  \Pi_{\varepsilon'} 
\ee
for a family of orthogonal projectors $\{\Pi_\varepsilon\}_\varepsilon$, where each pair $(\varepsilon,\varepsilon')$ with $\varepsilon\neq \varepsilon'$ can only appear once (for one $\omega$). In words, the operators $A_\omega$ describe jumps between subspaces labeled by $\varepsilon'$ and $\varepsilon$ appearing in $P_\omega$. A simple example, is where $\Pi_\epsilon$ defines the subspace associate to energy $\epsilon$ of the probe's Hamiltonian $H$  \cite{breuer2002theory}. Then, $P_\omega$ might collect all pairs of energies $(\varepsilon,\varepsilon')$ such that $\varepsilon'- \varepsilon =\omega$. We chose to write the jump operators in a more abstract form  so that the master equation~\eqref{eq: diss} can be applied very generally, in particular to the local  master equation~\cite{Hofer2017,Gonzlez2017,Scali2021},  partial secular approximations~\cite{Tscherbul2015,Farina2019,Cattaneo2019,Cattaneo2020,McCauley2020,trushechkin2021unified}, as well as reaction-coordinate mappings to deal with strong coupling \cite{Nazir2018}. 

The Hamiltonian $H_{{LS}}$ is the Lamb shift correction to the energy of the probe system induced by the interaction with the sample. 
$H_{LS}$ is typically small as compared to the energy of the probe and is very often neglected in the thermometry literature (see Ref. \cite{Higgins2013} for an exception). Nevertheless, it is generally a \emph{temperature dependent} correction $\dot{H}_{LS}\neq 0$ and may play an important role in low-temperature thermometry as we shall see.

\subsection{The master equation from a microscopic model}
    \label{sec: microscopic}
To vehicle some physical insight into Eq.~\eqref{eq: diss} let us give explicit expressions   for the microscopic derivation of the so-called global master equation \cite{breuer2002theory}. 
As the starting point consider the probe-sample coupling given by the global Hamiltonian in the form
\be
H_{\rm tot} = H + H_B + A\otimes B,
\label{eq:Htot}
\ee
where $H (H_B)$ is the Hamiltonian of the probe (bath or sample), and the interaction Hamiltonin has a form  $H_I= A \otimes B $ with Hermitian operators $A$ and $B$ acting on the probe  and sample respectively. It is convenient to diagonalize the probe Hamiltonian 
\be
H = \sum_{\varepsilon \in \t{Sp}(H)} \varepsilon\,  \Pi_\varepsilon,
\ee
where $\t{Sp}(H)$ is the spectrum of $H$,
defining the energy subspaces given by the set of orthogonal projectors $\{\Pi_\varepsilon\}$.

Assuming that the coupling between the probe and the sample is weak and using the Born-Markov and secular approximations\footnote{The secular approximation requires that any non-zero frequency $w=\varepsilon' -\varepsilon$ is larger than the interaction strength, see e.g. ~\cite{Tscherbul2015,Farina2019,Cattaneo2019,Cattaneo2020,McCauley2020,trushechkin2021unified}.}, one can derive the master equation for the probe with of the form of Eq.~\eqref{eq: diss} \cite{breuer2002theory}: 
\be\begin{split}
\label{eq: global master equation}
    \mathcal{D}(\rho) &= \sum_\omega \gamma_\omega \left( A_\omega \rho A_\omega^\dag- \frac{1}{2}\{ A_\omega^\dag A_\omega,\rho\} \right)\\
     H_{{LS}} &= \sum_\omega s_\omega \,  A_\omega^\dag A_\omega \\
     A_\omega &= \sum_{\varepsilon' -\varepsilon =\omega}  \Pi_\varepsilon A\,  \Pi_{\varepsilon'} 
\end{split}
\ee
The coefficeints $\gamma_\omega$ and $s_\omega$ are real and imaginaly parts of the correlation function of the reservoir $\Gamma_\omega = \frac{1}{2}\gamma_\omega+ \ii s_\omega = \int_0^\infty e^{\ii \omega t} \mean{B(0) B(t)} \dd t$ where $B(t)\equiv e^{-iH_B t} B e^{-iH_B t}$ are operators of the sample in the interaction picture.  One notes that Eq.~\eqref{eq: global master equation} is a particular case of Eqs.~(\ref{eq: diss}-\ref{eq: jump operators 2}).

In the following we consider the possibility of engineering the probe Hamiltonian $H$. In contrast, we treat the operator $A$ describing the coupling of the probe to the sample as being fixed.

\subsection{Fast quantum control}

On top of the thermalisation process, we assume that the probe and an additional auxilliary system of arbitrary dimension (not interacting with the sample) can be externally measured and controlled at will. The control capabilities are introduced in two steps. First, the  Hamiltonian $H$ (acting on the probe and the ancillae) can be engineered. In principle, we allow it to  vary in time as long as instantaneous thermalization at $t$ is well described by Eq.~\eqref{eq: diss} for some fixed $H(t)$ \footnote{This is well justified when the driving is sufficiently slow, for the so called adiabatic master equation \cite{Albash2012,Dann2018}, or if the driving is very fast, e.g. quenches, so that the bath does not have time to react. However, we will see that all protocols considered here, including optimal ones, do not require driving of $H$ when the QFI is the figure of merit. It is enough to optimise $H$ at the beginning of the process. }. Second, the dissipative evolution is being interjected with arbitrary control operations and measurements on the probe and the ancillae. Following~\cite{ffqc} this can be formally described  by dividing the total time $\tau$ in $m$ steps, $\dd t=\tau/m$ (in principle $m$ can be arbitrary large). The systems evolve accordingly to the master equation during each step  $e^{\dd t \cL}$ but in-between any two steps  an arbitrary gate, measurement or initialization can be applied. In fact because the dimension of the auxiliary system is kept free, all control operations can be implemented as unitary gates $U_k$. In particular, the information acquired with a measurement along the process  can mapped coherently onto some ancilla and used in the next steps, thus allowing for adaptive schemes. It is important to stress that, as above, we assume that the control operations do not hinder the approximations and time-scale separation used to derive the master equation~\eqref{eq: diss} in the first place. 

At any time $\tau$  the state of the probe and ancillae is given by a density matrix $\rho(\tau|T)$. The QFI of this state with respect to the temperature sets a fundamental limit on the precision of the whole scheme seen as a thermometer. Our goal is then to develop finite-time strategies that maximise the QFI of the channel 
\be\label{eq: tau channel}
\cE_{\tau|T } = e^{\dd t \cL}\cdot U_m\dots \circ  e^{\dd t \cL} \circ U_1.
\ee
Crucially, the control and presence of ancillae can not alter the coupling  $H_I = A \otimes B$ between  the probe and the sample. The intuition is that $H_I$ gives the bottleneck for how fast a change of temperature in the sample causes an effect on the probe, setting a limit on thermometry even in presence of arbitrary quantum control. 



\subsection{QFI bounds for control-assisted Markovian semi-group dynamics}

It is straightforward to verify that a thermalization of an infinitesimal duration $\dd t$ is given by a quantum channel $e^{\dd t \mathcal{L} }$ which admits a Kraus representation $\bm K =(K_0 \dots K_\omega \dots)$ composed of Kraus operators
\be\label{eq: Kraus }
\begin{split} 
    K_0 &= \id - \dd t \frac{1}{2} \sum_\omega \gamma_\omega  A_\omega^\dag A_\omega  - \ii\,  \dd t\, ( H +H_{LS})\\
    K_\omega &= \sqrt{ \dd t \gamma_\omega } A_\omega \qquad \forall\,  \omega,
\end{split}
\ee
 where the terms of order $\dd t ^{3/2}$ and higher are ignored. A Kraus representation of a quantum channel is of course not unique, any vector of operators $\tilde{\bm K} = u \bm K$ with a unitary gauge matrix $u$ describes the same channel\footnote{Note that the gauge matrix can explicitly depend on the parameter}. It has been shown in~\cite{ffqc,Rafal2017} that if there exist a Hermitian matrix $h=- \ii u^\dag \dot u$, such that 
\be\label{eq: diffusive}
\qquad \| \bm K^\dag ( \dot{\bm K} +\ii h \bm K )\| = O(\dd t^{3/2}), 
\ee
then the QFI of any channel $\cE_{\tau|T}$ of form of Eq.~\eqref{eq: tau channel} is upper bounded by
\be\label{eq: QFI upper}
\cF_{\cE_{\tau|T}} \leq 4 \frac{\tau}{\dd t} \| (\dot{\bm K}^\dag - \ii \bm K^\dag h)( \dot{\bm K} +\ii h \bm K )\|,
\ee
with $\|... \|$ the operator norm, which is given  by $\|A \| \equiv \max_j \lambda_j  $ where $\lambda_j$ are the eigenvalues of the Hermitian operator~$A$.  Note that if the master equation has an explicit time dependence, but Eq.~\eqref{eq: diffusive} is fulfilled at all times for some $h_t$, one can simply replace the bound with $\cF_{\cE_{\tau|T}} \leq 4 \int_0^\tau \| (\dot{\bm K}^\dag_t - \ii \bm K_t^\dag h_t)( \dot{\bm K}_t +\ii h_t \bm K_t )\|$,  this is analogous to the phase-estimation scenario with time-dependent noise discussed in the supplementary material of~\cite{Rafal2017}.

\section{Results}  
\label{sec: results}

We are now ready to present the main results of this article. We start by showing that the diffusive scaling $\cF_{\cE_{\tau|T}}\propto \tau$ induced by  Eq. \eqref{eq: QFI upper} applies to a wide class of probe-sample interactions.  We then derive upper bounds on $\cF_{\cE_{\tau|T}}$ for both fixed and engineered probe's Hamiltonian $H$,  and show how to saturate them by simple measurement strategies. Finally, we discuss the potential effect of the  Lamb shift Hamiltonian in thermometry.  
 
\subsection{The diffusive scaling of the QFI in thermometry}
\label{sec: diffusive?}

The first question to ask is whether ballistic scaling of QFI is possible in thermometry, i.e. whether there exists a $h$ satisfying Eq.~\eqref{eq: diffusive} as we just discussed. Following the argumentation of \cite{Rafal2017}, we derive the general condition for diffusive scaling of the QFI that can be found in the Appendix~\ref{app: deffusive}. 
A simpler but slightly more restrictive version of it reads
\be \label{eq: diffusive?}
\dot{H}_{LS}\in \,  \t{span}\left\{\id, \gamma_\omega A_\omega^\dag A_\omega \right \},
\ee
see Appendix~\ref{app: deffusive}. In words, if the derivative of the Lamb shift Hamiltonian $\dot H_{LS}$ is linearly dependent of identity $\id$ and the operators $\gamma_\omega A_\omega^\dag A_\omega$ for all $\omega$, then $\cF_{\cE_{\tau|T}}\propto \tau$.   It is worth noting that the condition~\eqref{eq: diffusive?} is also sufficient to enforce a diffusive scaling of the QFI if the Lamb shift Hamiltonian is block-digonal $H_{LS} =\sum_{\varepsilon} \Pi_\varepsilon H_{LS} \Pi_\varepsilon$, c.f. Appendix~\ref{app: deffusive}. 

In the case of our microscopic model introcuced in section \ref{sec: microscopic} the Lamb shift Hamiltonian takes the form $H_{LS} = \sum_\omega s_\omega A^\dag_\omega A_\omega$. Clearly, a sufficient condition for \eqref{eq: diffusive?} is that all the jump rates $\gamma_\omega$ are non-vanishing whenever the jump operator $A_\omega\neq 0$ and the derivative of the Lamb shift contribution $\dot{ s}_\omega\neq $ are non zero. We argue that this is usually the case. It is possible that some frequencies $\omega$ are missing in the spectrum of the bath, however in this case both $\gamma_\omega$ and $s_\omega$ are identically zero. If the frequency is present it typically gives rise
to the dissipative term $\gamma_\omega$, and not just a coherent temperature dependent Lamb shift $\dot s_\omega$. In such cases we can conclude that the QFI is bound to a diffusive scaling 
\be 
\cF_{\cE_{\tau|T}} \leq \t{const}\,  \tau.
\ee
Intuitively speaking, this scaling is a consequence of the inevitable presence of noise/dissipation in the process of temperature estimation, which prevents the ballistic scaling  $\cF_{\cE_{\tau|T}}\propto \tau^2$ commonly obtained in quantum-coherent evolutions (e.g. in noiseless phase interferometry \cite{DemkowiczDobrzaski2015}). 

On the other hand, it is interesting to ask the question whether one can engineer a more exotic reservoir in such a way that the condition for diffusive scaling of the QFI is not fulfilled.  For example, if the interaction in \eqref{eq:Htot} is replaced by the more general form $\sum_i A_i \otimes B_i$, then the diffusive scaling of the QFI is not guaranteed by the argument above.    
In this case,  ballistic scaling of the QFI could be achieved, possibly requiring  active control~\cite{zhou2018}.

 \subsection{Upper bound on QFI for a fixed $H$ (no Lamb shift)}
 
For clarity of the exposition, we start by ignoring the Lamb shift $H_{LS}$ term in the master equation in the following sections.
More precisely, we assume that the Lamb shift is constant around the true value of the bath temperature ($\dot s_\omega =0$ if $A_\omega^\dag A_\omega \neq 0$), so that it can be ignored. Indeed, any parameter independent term in the Hamiltonian can be cancelled via appropriate control operations. We will come back to the the effect of $H_{LS}$ later.

We start by considering a fixed probe Hamiltonian $H$, and discuss its engineering in the following section. In absence of Lamb shift Eq.~\eqref{eq: diffusive} can be satisfyed by setting $h=0$. This gives a particularly simple upper-bound on the QFI for thermalizaion. Straightforward application of Eq.~\eqref{eq: QFI upper} gives a very intuitive expression
\be\label{eq: QFI bound final main}
\cF_{\cE_{\tau|T}} \leq \tau \left \|\sum_w \frac{\dot \gamma_{w}^2}{\gamma_{w}} \,A_w^\dag A_w \right \|,
\ee
see Appendix~\ref{app: QFI bound}. 
Note that the bound only depends on the jump operators $A_\omega$ and the rates $\gamma_\omega$, and neither requires solving the master equation nor considering explicit control strategies. Remarkably, this bound can be saturated with a simple strategy as we now show.

\subsection{Saturation of the bound~\eqref{eq: QFI bound final main}  with a continuous measure-and-prepare scheme}

To save some space let us denote the norm appearing in the rhs of Eq.~\eqref{eq: QFI bound final main}  as $\mathds{O} = \left\|\sum_w \frac{\dot \gamma_{w}^2}{\gamma_{w}} \,A_w^\dag A_w \right\|$. For simplicity we start with the expression of the the jump operators $A_\omega$ in Eq.~\eqref{eq: global master equation}. It gives
\be\begin{split} 
\mathds{O} & 
= \left\|\sum_\omega \frac{\dot \gamma_\omega^2}{\gamma_\omega}
\sum_{\bar \varepsilon' -\bar \varepsilon =\omega}  \sum_{\varepsilon' -\varepsilon =\omega}  \Pi_{\bar \varepsilon'} A\,  \Pi_{\bar \varepsilon} \Pi_\varepsilon A\,  \Pi_{\varepsilon'}\right\|\\
& = \left\|\sum_\omega \frac{\dot \gamma_\omega^2}{\gamma_\omega} 
 \sum_{\varepsilon' -\varepsilon =\omega}  \Pi_{\varepsilon'} A\,  \Pi_{\varepsilon}  A\,  \Pi_{\varepsilon'}\right\|.
\end{split}
\ee
Here, we used $ \Pi_{\bar \varepsilon} \Pi_\varepsilon = \delta_{\bar \varepsilon \varepsilon} \Pi_\varepsilon$ and the fact that for a fixed $\omega$ and $\varepsilon$ there can only be one $\varepsilon'$  fulfilling $\varepsilon' -\varepsilon =\omega$. 
Since all the operators $\Pi_{\varepsilon'}O \Pi_{\varepsilon'}$ for different $\varepsilon'$ are orthogonal, we can simplify the operator norm to  
\be
\label{eq: first bound O}
\mathds{O} = \max_{\varepsilon'}  \left \|\Pi_{\varepsilon'} \left(\sum_{\varepsilon\neq \varepsilon'} \frac{\dot \gamma_{\varepsilon'-\varepsilon}^2}{\gamma_{\varepsilon'-\varepsilon}}\,  A \, \Pi_\varepsilon\, A \, \right) \Pi_{\varepsilon'} \right\|.
\ee
We denote  the energy which attains the maximum  by $\varepsilon_*$.

With the more general form of the jump operators $A_\omega = \sum_{(\varepsilon',\varepsilon) \in P_\omega}  \Pi_\varepsilon A_\omega  \Pi_{\varepsilon'}$ in Eq.~\eqref{eq: jump operators 2} the expression takes the form
\be
\label{eq: first bound O bis}
\mathds{O} = \max_{\varepsilon'}  \left \|\Pi_{\varepsilon'} \left( \sum_{\substack{\omega ,\varepsilon\\ (\varepsilon,\varepsilon')\in P_\omega}} \frac{\dot \gamma_{\omega}^2}{\gamma_{\omega}}\,  A_\omega^\dag \, \Pi_\varepsilon\, A_\omega \, \right) \Pi_{\varepsilon'} \right\|,
\ee
which follows from identical arguments, with the maximum attained for $\varepsilon'=\varepsilon_*$.

So far we have just rewritten Eq.~\eqref{eq: QFI bound final main}. To show that it can be attained, consider the following strategy. Prepare the probe in some state $\ket{\Psi_{\varepsilon_*}} = \Pi_{\varepsilon_*} \ket{\Psi_{\varepsilon_*}}$ in the $\varepsilon_*$-subspace. Let it evolve for an infinitesimal time $\dd t$ and measure
in which subspace $\Pi_\varepsilon$ it is (i.e. the energy of the probe). The probability that the probe jumps to another energy subspace labeled by $\varepsilon\neq \varepsilon_*$ is given by
\be\begin{split} 
    p_\varepsilon &= \tr \Pi_\varepsilon\,  e^{\mathcal{L}\dd t}\left( \prjct{\Psi_{\varepsilon_*}}\right) \\
    &=\dd t\,  \tr \Pi_\varepsilon  \sum_\omega \gamma_\omega A_\omega \prjct{\Psi_{\varepsilon_*}} A_\omega^\dag \\
    &= \dd t \, \gamma_\omega \bra{\Psi_{\varepsilon_*}} A_\omega^\dag \Pi_\varepsilon A_\omega \ket{\Psi_{\varepsilon_*}},
\end{split}
\ee
for $\omega=\omega(\varepsilon,\varepsilon_*)$ such that $(\varepsilon,\varepsilon_*)\in P_\omega$.
While the probability that it remains at $\varepsilon_*$ is simply $p_{\varepsilon_*} =1- \sum_{\varepsilon\neq \varepsilon_*} p_\varepsilon$. Computing the Fisher information for this measurement after $\dd t$  we find that in the leading order
\be 
\cI_{\dd t} = \sum_\varepsilon \frac{(\dot p_\varepsilon)^2}{p_\varepsilon} = \dd t \bra{\Psi_{\varepsilon_*}} \sum_{\substack{\omega ,\varepsilon\\ (\varepsilon,\varepsilon_*)\in P_\omega}} \frac{\dot \gamma_{\omega}^2}{\gamma_{\omega}}\,  A_\omega^\dag \, \Pi_\varepsilon\, A_\omega \ket{\Psi_{\varepsilon_*}}
\ee
By maximizing this expression over all states $\ket{\Psi_{\varepsilon_*}}$ we find the optimal Fisher information $\cI_{\dd t} = \dd t\, \left\| \Pi_{\varepsilon_*} \sum_{\omega ,\varepsilon} \frac{\dot {\gamma}_{\omega}^2}{\gamma_\omega}\,  A_\omega^\dag \, \Pi_\varepsilon \, A_\omega \Pi_{\varepsilon_*} \right\|$ by definition of the operator norm.  After the measurement the probe is reinitialized to the optimal state $\ket{\Psi_{\varepsilon_*}}$, and the same procedure is repeated. By additivity, the overall Fisher information of this scheme reads
\be
\cI_\tau = \int_0^\tau \cI_{\dd t} = \tau \left \| \Pi_{\varepsilon_*} \!\!\! \sum_{\substack{\omega ,\varepsilon\\ (\varepsilon,\varepsilon_*)\in P_\omega}} \frac{\dot \gamma_{\omega}^2}{\gamma_{\omega}}\,  A_\omega^\dag \, \Pi_\varepsilon\, A_\omega  \, \Pi_{\varepsilon_*} \right \|
\label{eq:tight}
\ee
which coincides with the upper-bound as expressed in the form of Eq.~\eqref{eq: first bound O bis} (and also Eq.~\eqref{eq: first bound O} as a particular case).

This shows that the upper-bound \eqref{eq: QFI bound final main} is indeed saturated with a simple continuous measure-and-prepare strategy, where the jumps (the energy) of the probe are constantly monitored and the probe is actively reinitialized to the initial state. The optimality of fast-measurements for qubit probes was already noted numerically in Ref. \cite{Correa_2015} (see also \cite{Pasquale_2017,Mehboudi_2019rev}), and we generalised these results to general probes and arbitrary control-assisted/adaptive schemes.  In contrast to such previous results, it is worth stressing that our approach neither requires solving the Lindblad dynamics nor considering specific strategies, instead the optimal protocol can be inferred directly from simple calculation of~\eqref{eq: QFI bound final main} and~\eqref{eq:upboundO}. This enables us to study optimal protocols for more complex interactions and probes, as we will illustrate in Sec.~\ref{Sec: Case Studies}.

 \subsection{Engineering of the optimal spectrum for $H$}
 \label{sec: optimal H}
 
The bound \eqref{eq: QFI bound final main} is saturable and holds independently of the unitary control operations $U_k$ applied during the sensing process. Nevertheless, in the context of the microscopic model in Sec. \ref{sec: microscopic} it does depend on the probe Hamiltonian $H$ via the jump operators $A_\omega$ and the corresponding rates. The goal of this section is to find the Hamiltonian maximizing the QFI of our thermometer, given the form of the observable $A$ coupling the probe to the sample, its temperature $T$ and the spectral dependence of the rates $\gamma_\omega$ and $\dot{\gamma}_{\omega}$. Recall that we allow the usage of auxiliary systems of arbitrary dimension. Hence, we consider Hamiltonians $H =\sum_\varepsilon \varepsilon \Pi_\varepsilon$ acting on the probe and the auxiliary system, while only the probe couples to the sample $A = A_\t{probe}\otimes\id_\t{ancillae}$ (this is implicit in the following discussion).

To optimize over all Hamiltonians $H$ one can vary its spectrum $(\varepsilon_1,\varepsilon_2,\dots)$ and the subspaces $\sum_\varepsilon \Pi_\varepsilon =\id$ independently. To find the best Hamiltonian we have to maximize the quantity $\mathds{O} = \left\|\sum_w \frac{\dot \gamma_{w}^2}{\gamma_{w}} \,A_w^\dag A_w \right\|$. Starting with any Hamiltonain let us label the energy subspace that attains the maximum in Eq.~\eqref{eq: first bound O} with $\varepsilon_*$, without loss of generality we can shift all the energies such that $\varepsilon_* = 0$ ($\Pi_{\varepsilon_*}\to \Pi_0$). The sum inside the norm in Eq.~\eqref{eq: first bound O} involves positive operators $\Pi_0 A \Pi_{\varepsilon} A \Pi_0 = (\Pi_0 A \Pi_{\varepsilon})(\Pi_0 A \Pi_{\varepsilon})^\dag$ and positive scalar coefficients $(\dot \gamma_{-\varepsilon}^2/\gamma_{-\varepsilon})$. Therefore, for fixed subspaces $\Pi_\varepsilon$ the expression $\mathds{O}$ is maximized when all the scalar coefficients are maximized. But for any bath temperature $T$ there is indeed an optimal energy difference $\delta_T$ maximizing the coefficients
\be 
f_T =\max_\varepsilon \frac{\dot \gamma_{-\varepsilon}^2}{\gamma_{-\varepsilon}},
\qquad
\delta_T = -\t{argmax}_\varepsilon \frac{\dot \gamma_{-\varepsilon}^2}{\gamma_{-\varepsilon}}.
\ee
Hence, for a fixed $\Pi_0$ it is optimal to set the energies of all the other levels to the same value $\varepsilon = \delta_T$. Using $\sum_{\varepsilon \neq 0} \Pi_\varepsilon = \id -\Pi_0$ we get for the optimal choice of energies
\be 
\mathds{O} =  f_T\,   \| \Pi_0 A (\id-\Pi_0) A \Pi_0 \|.
\ee 
For a fixed $\Pi_0$ the norm is attained by some state $\ket{\Psi}= \Pi_0 \ket{\Psi}$. Denoting $\Pi_{0\setminus \Psi} = \Pi_0 - \prjct{\Psi}$ we obtain
\be
\begin{split}
     \| \Pi_0 A (\id-\Pi_0) A \Pi_0 \| 
     &= \bra{\Psi} A (\id-\prjct{\Psi} -\Pi_{0\setminus \Psi}) A \ket{\Psi} \\
     &\leq \bra{\Psi} A (\id-\prjct{\Psi}) A \ket{\Psi}.
\end{split}
\ee
But changing the Hamiltonian to one with $\Pi_0 =\prjct{\Psi}$ saturates this inequality. Thus, an optimal Hamiltonian is necessarily of this form $H = \delta_T (\id-\prjct{\Psi})$, which is fully degenerate except for one energy eigenstate (coincidentally, this same structure is optimal for equilibrium thermometry \cite{Correa_2015}). It remains to find the optimal state~$\ket{\Psi}$. For a rank one projector $\Pi_0=\prjct{\Psi}$ the expression $\| \Pi_0 A (\id-\Pi_0) A \Pi_0 \| = \bra{\Psi}A^2 \ket{\Psi} -  \bra{\Psi}A \ket{\Psi}^2$ is simply the variance of $A$ with respect to the state $\ket{\Psi_*}$. The variance of an operator is maximized on the state, which is an equal superposition  
\be\label{eq: GHZ}
\ket{\Psi_{\Delta(A)}} =\frac{1}{\sqrt 2}(\ket{a_\t{max}}+\ket{a_\t{min}})
\ee
of two extremal eigenstates of $A\ket{a_\t{max(min)}} = \lambda_\t{max(min)}\ket{a_\t{max(min)}}$. As a result we obtain the simple and general bound
\be\label{eq:upboundO}
\cF_{\cE_{\tau|T}} \leq \tau \left(\max_\omega \frac{\dot \gamma_{\omega}^2}{\gamma_{\omega}}\right) \left(\frac{\Delta(A)}{2}\right)^2 ,
\ee
where $\Delta(A) = \lambda_\t{max} - \lambda_\t{min}$ is the spectral gap of the operator $A$ (the difference between its  maximal and minimal eigenvalues).

In summary, we have shown that the QFI of our thermometer is maximized for the almost degenerate Hamiltonian 
\be\label{eq: optimal H}
H = \delta_T(\id-\prjct{\Psi_{\Delta(A)}}),
\ee
with the unique nondegenerate energy level $\ket{\Psi_{\Delta(A)}}$ of the form in Eq.~\eqref{eq: GHZ}. This bound allows us to formalize the original intuition -- the form of the operator $A$ coupling the probe to the sample does set a general  ``speed limit" on thermometry. Recall that the bound can be attained with the continuous measure-and-prepare strategy, where the probe is  reinitialized  to the state $\ket{\Psi_{\Delta(A)}}$. It is worth noting that this strategy does not require an auxiliary system.

\subsection{Autonomous implementation of the fast measure-and-prepare strategy}
\label{sec: autonomous}

\begin{figure}
    \centering
    \includegraphics[width=0.9 \columnwidth]{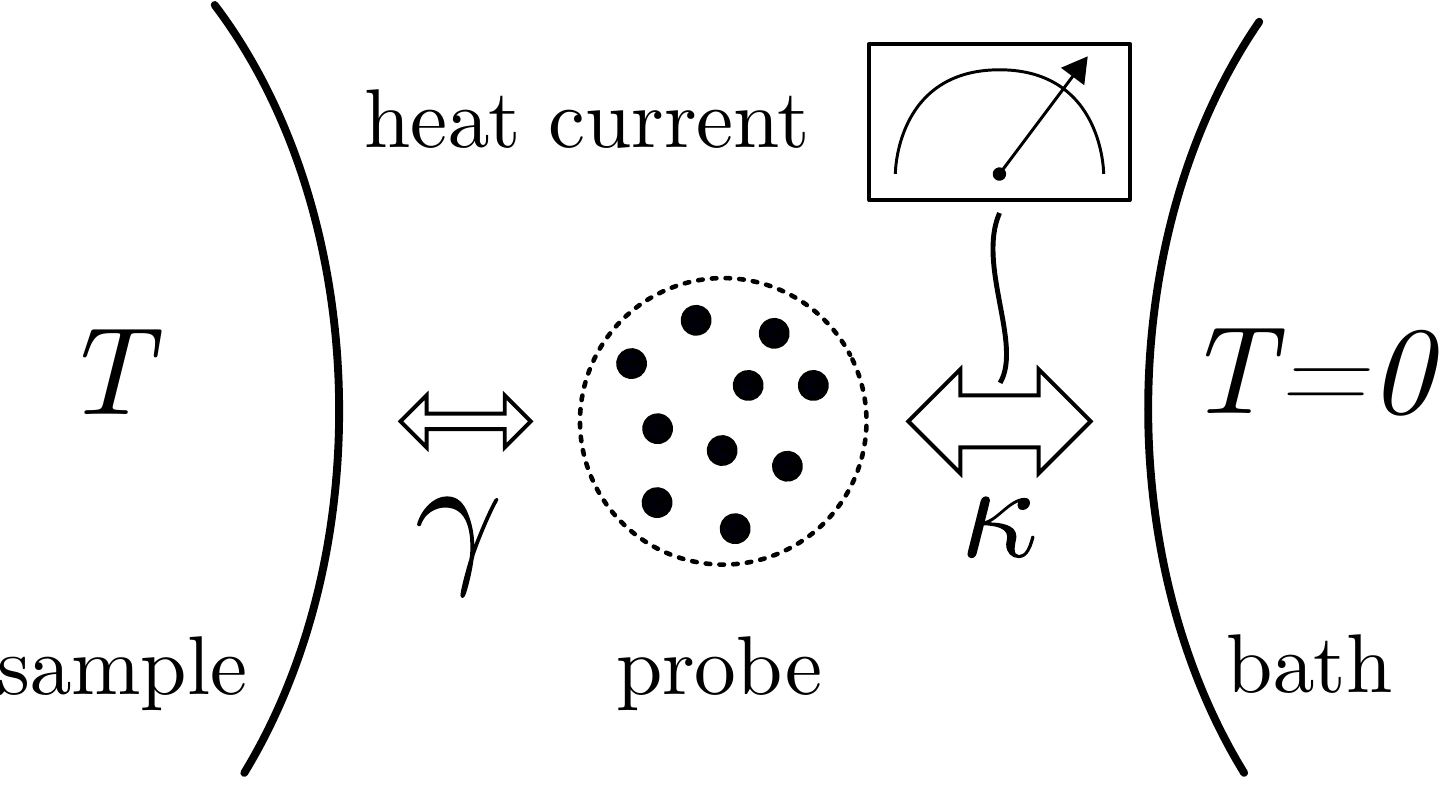}
    \caption{A practical autonomous thermometer scheme. In the limit where the interaction of the probe with the cold bath is much stronger then its interaction with the sample the probe is constantly reset to its ground state. Continuous measurements of the heat current between the probe and the cold bath allow one to monitor the energy jumps in the probe induced by the sample. Generally, this setup implements the continuous measure and prepare strategy, where the probe is reset to its ground state. In particular, for collective thermalisation (stemming from probe-sample interaction of the form $H_I=J_x\otimes B$) and optimized two-body probe Hamiltonian of the form $H = \frac{\omega}{2}J_z +  b J_z^2$ we show in Sec. \ref{sec: Collective thermalisation} that the strategy comes very close to saturating the upper-bound of Eq.\eqref{eq:upboundO}.}
    \label{fig:auto}
\end{figure}

In order to exploit these ideas in practice, a possible autonomous implementation is depicted in Fig.~\ref{fig:auto}. We consider a scenario where the probe is simultaneously coupled to the sample and to a zero temperature  bath ($T=0$). If $\kappa$ and $\gamma$ are parameters describing the coupling strength between the probe and each bath, $\kappa$ for the zero-temperature bath and $\gamma$ for the sample, we consider the regime $\kappa \gg \gamma$. 
This ensures that, every time the probe gains an excitation due to the interaction with the sample, the cold bath quickly reinitialises the probe back to its ground state. By monitoring the heat current between the probe and the cold bath  at the single-photon level, one is able to keep track of the energy jumps induced on the probe by the sample (the same approach is used in quantum thermodynamics to measure heat and work statistics, see e.g.~\cite{Pekola2013,Menczel2020}).
This provides a simple way of performing the fast measure-and-prepare strategy introduced earlier, in which the cold bath simultaneously prepares and measures the probe. It is worth mentioning that this type of environment monitoring to measure single-photon emissions  is already being implemented in superconducting quantum circuits~\cite{Pekola2013,Gasparinetti2015,Govenius2016,Karimi2020}. 

To saturate the upper-bound~\eqref{eq:upboundO} one needs to implement the optimal probe Hamiltonian of the form given in Eq.~\eqref{eq: optimal H}. This might not be an easy task in practice. Nevertheless, simpler Hamiltonians might already come close to the upper-bound. This will be illustrated in  Sec.~\ref{sec: Collective thermalisation} for  collective two-body Hamiltonians which are implemented in spin squeezing experiments~\cite{Treutlein}.

\subsection{Effect of the Lamb shift}

We have derived very general and attainable bounds on the control-assisted thermometry both in the case of a fixed probe Hamiltonian $H$ in \eqref{eq: QFI bound final main} as well as for an optimized $H$ in \eqref{eq:upboundO}. In doing so we have however assumed that the  Lamb shift term in the master equation~\eqref{eq: diss} can  be ignored\footnote{This assumption is common in the literature on quantum thermometry.}. Let us now   assume the presence of a Lamb shift satisfying the condition~\eqref{eq: diffusive?} and include it in the analysis.

As shown in Sec.~\ref{sec: diffusive?}, the time scaling of the Fisher information is still diffusive $\cF_{\cE_t|T} \propto \tau$ in presence of a temperature-dependent Lamb shift if the condition~\eqref{eq: diffusive?} holds. In this case, the derivative of the Lamb shift Hamiltonian can be expressed as
\be\label{eq: HLS decomposition}
\dot{H}_{LS} = h_{\id} \, \id + \sum_{\omega} h_\omega A^\dag_\omega A_\omega,
\ee
with some real coefficeints $h_\id$ and $h_\omega$ such that $h_\omega=0$ if $\gamma_\omega=0$. Then, we show in the appendix~\ref{app: QFI bound} that the bounds for a fixed $H$~\eqref{eq: QFI bound final main} and for the optimal $H$ ~\eqref{eq:upboundO} are modified to the form
\be\label{eq: QFI bound lamb general}\begin{split}
\cF_{\cE_\tau|T} &\leq \tau \left \|\sum_\omega \frac{\dot \gamma_\omega^2+ 4 h_\omega^2}{\gamma_\omega} \,A_\omega^\dag A_\omega \right\|,\\
\cF_{\cE_{\tau|T}} &\leq \tau \left(\max_\omega \sum_\omega \frac{\dot \gamma_\omega^2+ 4 h_\omega^2}{\gamma_\omega}\right) \left(\frac{\Delta(A)}{2}\right)^2.
\end{split}
\ee
Here, to get the second inequality one simply repeats the derivation of section~\ref{sec: optimal H} for the new scalar functions $\left(\frac{\dot \gamma_\omega^2}{\gamma_\omega}\right) \mapsto \left(\frac{\dot \gamma_\omega^2+ 4 h_\omega^2}{\gamma_\omega}\right)$.

In the case of the microscopic model, a particularly simple choice of the coefficients $h_\id$ and $h_\omega$ in Eq.~\eqref{eq: HLS decomposition} is offered by the original expression for the Lamb shift 
\be
\dot{H}_{LS} =\sum_{\omega} {\dot s}_\omega A^\dag_\omega A_\omega
\ee
assuming  that $\gamma_\omega\neq 0$ if ${\dot s}_\omega \neq 0$. This choice gives rise to simple upper bounds 
\be\label{eq: QFI bound lamb}\begin{split}
\cF_{\cE_\tau|T} &\leq \tau \left \|\sum_\omega \frac{\dot \gamma_\omega^2+ 4 \dot s_\omega^2}{\gamma_\omega} \,A_\omega^\dag A_\omega \right\|,\\
\cF_{\cE_{\tau|T}} &\leq \tau \left(\max_\omega \sum_\omega \frac{\dot \gamma_\omega^2+ 4 \dot s_\omega^2}{\gamma_\omega}\right) \left(\frac{\Delta(A)}{2}\right)^2.
\end{split}
\ee
that  are directly expressible with quantities appearing in the master equation~\eqref{eq: global master equation}. The disadvantage is that they might diverge in the low-temperature limit where the jump rates $\gamma_\omega$ can  become negligible compared to $\dot{s}_w$, as we will see in the next section.  In this case it is better to use the more general bounds of Eqs.~\eqref{eq: HLS decomposition} and~\eqref{eq: QFI bound lamb general} with an appropriate choice of $h_w$.

The main difference between Eq. \eqref{eq: QFI bound lamb general}  and ~\eqref{eq: QFI bound final main}  concerns the attainability of these bounds. First, it is important to note that the Fisher information of the fast measure-and-prepare strategy is not affected by a Lamb shift. Indeed, due to the Zeno effect it does not affect the measurement probabilities after an infinitesimal time $\dd t$. Therefore, the values on the rhs of the inequalities \eqref{eq: QFI bound final main} and \eqref{eq:upboundO} can still be attained in presence of a Lamb shift.
However, these values are not tight with the new upper-bounds. Furthermore, the Zeno effect also implies that to approach the new bounds one has to move away from continuous measurement schemes which are insensitive to Lamb shift.
Even if the bounds~\eqref{eq: QFI bound lamb general} and \eqref{eq: QFI bound lamb} are not necessarily attainable by any measurement strategy, it is important to stress that the presence of the Lamb shift can only improve the estimation precision. This will be illustrated in the following example, where we will see that the presence of Lamb shift can enable dramatic enhancements in the low-temperature regime. 



\section{Case studies}
\label{Sec: Case Studies}

In this section we apply our general results  to two pedagogical and experimentally relevant examples. First, we consider a qubit probe with an explicit model of the sample as a collection of bosonic modes. This allows us to obtain explicit expressions for the jump rates in terms of sample properties, and discuss different temperature regimes. Second, we consider the case of an $N$-qubit probe with a two-body Hamiltonian and coupled collectively to the sample. Here, we discuss the scaling of the thermometer precision with $N$.

\subsection{Qubit probe}
\label{sec: qubit probe}
 We consider a qubit probe weakly interacting with a bosonic sample (bath).  This is a common scenario in thermometry experiments at the quantum scale, such as in the temperature measurement of cold atomic ensembles  \cite{Recati_2005,Sabin_2014}, nanoresonators \cite{Brunelli_2011,Brunelli_2012}, and black body radiation thermometry  \cite{Norrgard2021} through  two-level probes.

We start from  the Hamiltonian
\begin{align}
    H_{\rm tot}=&\frac{w}{2}\sigma_z +\sigma_x \int_0^\Omega \sqrt{J(x)} \left(b_x +b_x^{\dagger}\right) {\rm d}x 
    \nonumber\\
   & +
    \int_0^\Omega  x b_x^{\dagger}b_x  {\rm d}x,
\end{align}
where $\omega$ is the frequency of the probe, $b_x$ ($b_x^\dagger$) are annihilation (creation) operators of a bosonic mode at frequency $x$,  $J(x)$ is the spectral density of the sample or bath, and $\Omega$ is a cut-off frequency.  This is a particular case of the model discussed in Sec.~\ref{sec: microscopic} with $H= \frac{w}{2}\sigma_z$, $A= \sigma_x$, $B =\int_0^\Omega \sqrt{J(x)} \left(b_x +b_x^{\dagger}\right) {\rm d}x$ and $H_B =\int_0^\Omega  x b_x^{\dagger}b_x  {\rm d}x$. Following the standard derivation of the master equation in the weak coupling regime through the Born-Markov and secular approximations \cite{breuer2002theory}, we obtain the Lindblad equation \eqref{eq: diss} where 
\be
A_w =\sigma_-\qquad A_{-w} =\sigma_+
\ee
and  the rates  $ \gamma_w =2 \pi  J(w) (1+N(w)) $, $ \gamma_{-w} = 2 \pi J(w) N(w) < \gamma_w$, with $N(w) = (e^{\beta w} - 1)^{-1}$ the Bose-Einstein distribution. Furthermore, the Lamb shift takes the form $H_{LS}=s_w A^{\dagger}_w A_w + s_{-w} A^{\dagger}_{-w} A_{-w},$ with
$s_w =-( \Delta_T + \Delta)$
, $s_{-w}= \Delta_T$
where $\Delta_T, \Delta$ are given by (see e.g. \cite{McCauley2020}):
\begin{align}
   & \Delta=  \mathbb{P} \left[\int_0^\Omega \frac{J(x)}{x-w} {\rm d}x \right]
    \nonumber\\ 
    &\Delta_T=  \mathbb{P} \left[\int_0^\Omega \frac{J(x) N(x)}{x-w} {\rm d}x \right].
\end{align}
Here $\mathbb{P}$ denotes the Cauchy principle value of the integral. Note that all the temperature dependence is encoded in $N(w)$, which enters into $\gamma_w$, $\gamma_{-w}$ and $H_{LS}$. This implies $\dot s_{-w}= -\dot s_w$ with
\be
\label{eq: derivative LS}
\dot{H}_{LS} = - \dot s_w\,  \sigma_z,
\ee
and $\dot \gamma_{-w} = \dot \gamma_{w}$. 
For concreteness, in what follows we assume a spectral density of the form 
\begin{align}
    J(w) =g w^{\alpha}, \hspace{8mm} w\leq \Omega
\end{align}
where $\alpha$ determines the Ohmicity of the sample ($\alpha=1$ for an Ohmic bath).

\subsubsection{Upper bounds on the QFI}

In order to compute Eq. \eqref{eq: QFI bound lamb general}, we first note that the condition~(\ref{eq: diffusive?},\ref{eq: HLS decomposition})
\be
\dot{H}_{LS} =- \dot s_w\,  \sigma_z = h_\id \id + h_{w}  \prjct{1}+   h_{-w}  \prjct{0}
\ee
implies
\be
\begin{cases}
h_w = \dot s_w - x \\
h_{-w} = -\dot s_{w} - x,
\end{cases}
\ee
leaving a free parameter $x=h_\id$. The bound~~\eqref{eq: QFI bound lamb general} then takes the form
\be\label{eq: bound qubit general}
\cF_{\cE_\tau|T}  \leq \tau \min_x \left \|
\begin{pmatrix}
\frac{\dot{\gamma}_{w}^2+ 4 (\dot{s}_w +x)^2}{\gamma_{-w}} & \\
 &  \frac{\dot{\gamma}_w^2+ 4 (\dot{s}_w - x)^2}{\gamma_w}
\end{pmatrix}
\right\|.
\ee
 In the appendix we derive the closed form expression of the bound optimized with respect to $x$. The resulting expression is a little lengthy, but two limiting regimes are simple to understand. In the low temperature limit $(T\to 0)$ the rate  $\gamma_{-w}$ goes to zero faster than $\dot s_w^2$. To cancel the diverging term $\frac{(\dot s_w +x)^2}{\gamma_{-w}}$ we then choose $x=-\dot s_w$ so that that the bound reads 
\be\label{eq: B T->0}
\cF_{\cE_\tau|T}  \leq \tau \frac{\dot \gamma_w^2 + 16 \dot s_w^2}{\gamma_w}.
\ee 
In the high temperature limit $(T\to \infty)$ the values $\gamma_w \approx \gamma_{-w}$ become relatively close. Then a good choice is to set $x=0$, yielding the bound 
\be\label{eq: B T->infty}
\cF_{\cE_\tau|T}  \leq \tau \frac{\dot \gamma_w^2 + 4 \dot s_w^2}{\gamma_{-w}}.
\ee  Of course both bounds~(\ref{eq: B T->0},\ref{eq: B T->infty}) remain valid for all $T$, but are the tightest in the respective temperature limits.

\begin{figure}
\vspace{3mm}
		\centering
		\includegraphics[width=1.\linewidth]{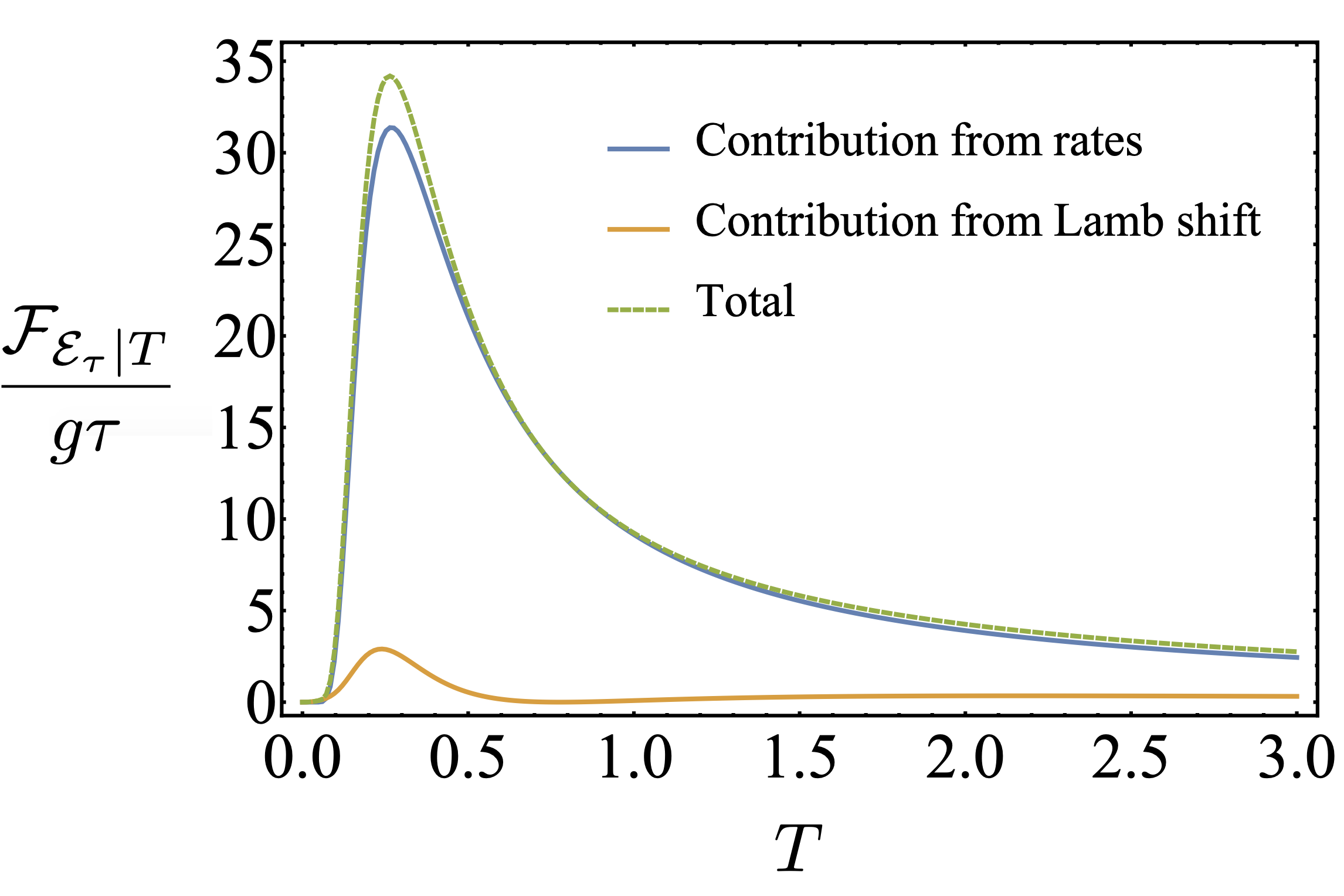} 
		\includegraphics[width=1.\linewidth]{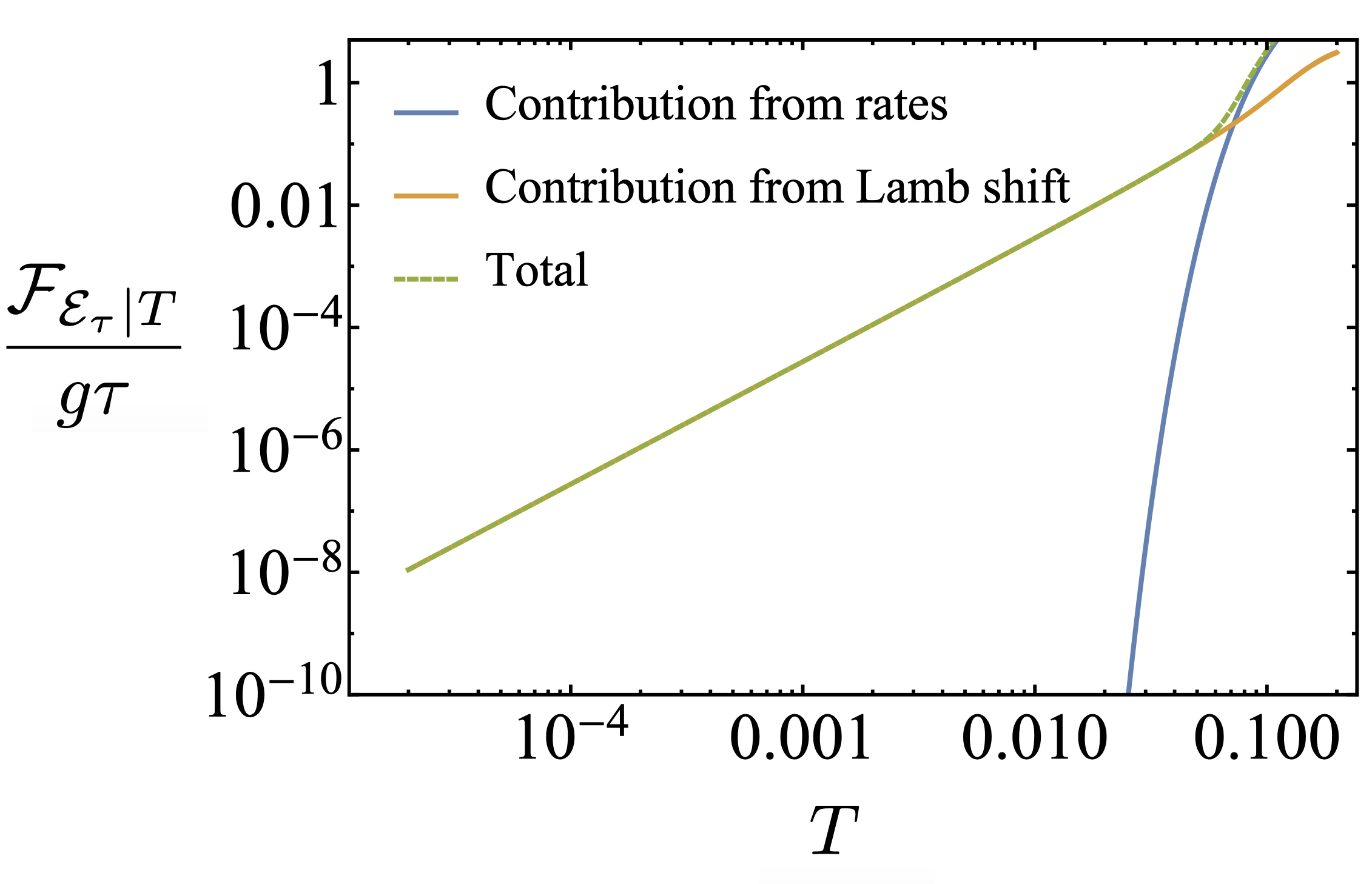} 
	    \caption{Plot of the upper bounds of $F$ in Eq.~\eqref{eq: bound qubit general} (in green). The different lines distinguish the contributions coming from either $\dot{\gamma}_w$ (rates, in blue) or $\dot{s}_w$ (Lamb shift, in orange).  For both Figures, we take $w=1$ and cutoff frequency $\Omega=5$. Figure  (a) shows the regime $T/w =\mathcal{O}(1)$, where the contribution from the rates is the most relevant one, whereas Fig. (b) focuses on the low-temperature regime $T/w \ll 1$, where the Lamb shit contribution highly dominates.  }
	\label{fig: qubit probe}
	\end{figure}

In Fig.~\ref{fig: qubit probe} we plot the upper bound \eqref{eq: bound qubit general} for an Ohmic sample (similar results are obtained for different $\alpha$). In the temperature range of  Fig.~\ref{fig: qubit probe}(a) we observe that most of the contribution to (the upper bound of) $\mathcal{F}$ arises from the rates, whereas the Lamb shift provides a small but non-negligible contribution. Interestingly, a completely different picture appears in the low-temperature regime $T\ll w$ of Fig.~\ref{fig: qubit probe}(b). In this case, it is well known that thermometry becomes exponentially hard: $\mathcal{F}\propto e^{-w/T}$~\cite{Paris_2015,Correa_2017,Hovhannisyan_2018}, as it can be inferred from  $N(w)=\mathcal{O}(e^{-w/T})$.    Remarkably, this is no longer the case if the Lamb shift contribution is not neglected. We find that $\Delta_T$ at low temperatures takes the form:
\begin{align}\label{eq:low T regime}
    \Delta_T \approx J(w) b_{\alpha} \left(\frac{T}{w} \right)^{1+\alpha}, \hspace{10mm} T\ll w,
\end{align}
where we have defined $b_{\alpha} \equiv  f_{\alpha} \Gamma(\alpha+1) $ with $\Gamma(\alpha)$  the Gamma function  and $f_{\alpha}=\mathcal{O}(1)$. From   \eqref{eq: B T->0} using $\gamma_w \approx 2\pi J(w) $  we obtain the upper bound
\begin{align}
   \cF_{\cE_{\tau|T}}\leq \tau \frac{8 J(w)(1+\alpha)^2 b_\alpha^2}{\pi w^2} \left(\frac{T}{w} \right)^{2\alpha } \quad \t{for}\quad T\ll w. 
\end{align}
Notably, for a flat spectral density ($\alpha=0$), the QFI can in principle remain constant with $T\rightarrow 0$, which is not in contradiction with the fundamental bounds derived in Ref. \cite{Potts_2019,Jrgensen2020}. In Sec.~\ref{subsec : Qubit scheme} we will discuss the saturability of this bound with explicit measurement schemes. 

\subsubsection{Optimal finite-time thermometry through a qubit probe}

We first consider the scenario where control over $w$ is  possible, so that the probe can be engineered to perform at the maximum QFI, as in Fig.~\ref{fig: qubit probe} (a). In this case, the Lamb shift plays a secondary role, and we  neglect it in this subsection\footnote{Note that accounting for the Lamb shift can only make the protocol better. At the same time, the fast measure-and-prepare strategy is completely insensitive to the presence of a Lamb shift. Hence, exploiting the Lamb shift requires finite times between subsequent measurements of the probe.}.
In this case, we can use the bound of Eq.~\eqref{eq: QFI bound final main}, which becomes $ \cF_{\cE_{\tau|T}}\leq \tau \frac{\dot \gamma_w^2}{\gamma_{-w}}$ and takes the explicit form
\begin{align}
\label{eq:boundqubit}
    \cF_{\cE_{\tau|T}}\leq \tau \gamma N(w)^3 e^{2w\beta } w^{2+\alpha} \frac{1}{T^4}.
\end{align} 
As discussed above, it can  be saturated with the fast measurement-and-prepare strategy where the probe is continuously measured and reinitialized in the ground state. The optimal frequency $w$, which depends on both $\beta$ and $\alpha$,  can be found numerically by maximising the right hand side of the inequality~\eqref{eq:boundqubit}, saturating \eqref{eq:upboundO} for this set-up. 
Setting $\alpha=3$ in \eqref{eq:boundqubit}, we recover the results of Ref.~\cite{Correa_2015}, where the performance of the ground-state  was compared to other suboptimal states (see in particular Fig. 2 of~\cite{Correa_2015}) and to strategies where the probe's state is not reinitialised \cite{Pasquale_2017,Mehboudi_2019rev}. 

\subsubsection{Optimal finite-time estimation of bath Ohmicity through a qubit probe}
\label{sec: Ohmicity}

One strength of our framework is that it allows one to  obtain similar results beyond temperature estimation. The determination of spectral properties of mesoscopic environments is a highly  non-trivial task, and realistic samples can often become  non-Ohmic~\cite{Grblacher2015,Lampo2017}.  Inspired by such results, we consider the estimation of the Ohmicity $\alpha\geq 0$ of a bosonic environment with rates $ \gamma_w = \gamma w^\alpha N(w) $,  $\gamma_{-w} = \gamma w^\alpha (1+N(w)) $ and $N(w) = (e^{\beta w} - 1)^{-1}$. In this case, the bound of Eq.~\eqref{eq: QFI bound final main} takes the simple form
\begin{align}
\label{eq:boundqubitII}
    \cF_{\cE_{\tau|\alpha}}\leq \tau (\ln w)^2 \gamma_{-w}.
\end{align} 
Again, this bound can be saturated by the fast measurement-and-prepare strategy.

\subsubsection{Low temperature regime: Lamb shift thermometry 
}
\label{subsec : Qubit scheme}

We have already noticed the potential importance of the Lamb shift contribution to the bound in the low-temperature regime through the upper bound shown in Fig. \ref{fig: qubit probe} (b). This is arguably the most challenging regime for thermometry, and several strategies for equilibrium thermometry based on strong sample-probe coupling have been suggested to avoid the exponential decaying  QFI \cite{Correa_2017,Hovhannisyan_2018,Potts_2019,Jrgensen2020}. 
We now sketch a new non-equilibrium approach that works in the weak coupling regime, and instead exploits the temperature dependence of the Lamb shift via a coherent evolution of the probe.

We start by explicitly solving the qubit master equation. In the limit $T\ll w$ the rate $\gamma_{-w}\to 0$, so that the jumps induced by the operator $A_{-w} = \sigma_+$ can be neglected. Furthermore, in this regime $\dot \gamma_w \ll \dot s_w$  as discussed around Eq.~\eqref{eq:low T} regime, and the contribution of $\dot \gamma_w $ to the QFI can also be neglected. The qubit master equation then becomes
\be\label{eq: qubit probe ME}\begin{split}
\frac{\dd}{\dd t} \rho &= -\ii \big[\frac{1}{2} w \, \sigma_z+H_{LS},\rho\big] \\&+\gamma_{w}\left( \sigma_- \rho\, \sigma_+ -\frac{1}{2}\big\{\prjct{1},\rho\big\}\right).
\end{split}
\ee

The Lamb shift contribution to the Hamiltonian reads $H_{LS} =
\begin{pmatrix} \Delta_T +\frac{1}{2}\Delta& \\
& -\Delta_T-\frac{1}{2}\Delta \end{pmatrix}$, where we shifted the energy scale to $\tr H_{LS}=0$. The solution of the master equation is straightforward, and can be found in  Appendix~\ref{app: ME qubit probe}. From Eq.~\eqref{eq: qubit probe ME} it is clear that we are dealing here with the problem of frequency estimation in presence of spontaneous decay noise ($T_1$-relaxation).

We consider three different sensing strategies, described in appendix~\ref{app: ME qubit probe} in detail. The simplest strategy (i) consists in preparing the qubit in the state~$\ket{\Psi_0}=\sqrt{1-a}\ket{0} + \sqrt{a} \ket{1}$, letting it evolve for some finite time~$t$, after which the qubit is measured and the procedure is repeated. For the second strategy (ii) we introduce an auxiliary qubit, which is not in contact with the sample. The probe and auxiliary qubits are repeatedly prepared in the state $\ket{\Psi_0}=\sqrt{1-a} \ket{00} + \sqrt{a}\ket{11}$, evolve freely, and are measured after some time $t$. The entanglement between the two qubits allows the final measurement to know if a jump $\sigma_-$ occurred during the evolution (such a jump projects the system to the state $\ket{01}$ changing the parity). The third strategy (iii) requires some continuous control on top of entanglement with the auxiliary qubit. Here, the qubits are again prepared in the state of the form $\ket{\Psi_0}=\sqrt{1-a} \ket{00} + \sqrt{a}\ket{11}$, but the parity of the pair is monitored continuously, with repeating the measurement described by the POVM $\{\Pi_\checkmark= \prjct{00}+\prjct{11}, \Pi_{\cross}=\id-\Pi_\checkmark\}$ as often as possible. If the error outcome "$\cross$" occurs the qubit pair is reinitialised. Otherwise, if no errors are observed for some time $t$ the system is measured and reinitialised. One notes that the coherent part of the dynamics $-\ii [\sigma_z\otimes\id, \rho]$ does note change the parity of the qubit pair and is unperturbed by the measurements $\{\Pi_\checkmark, \Pi_{\cross}\}$. The strategy (i) describes the usual Ramsey interferometer. The strategies (ii) and (iii) build on (i) by implementing the error-detection schemes considered in \cite{ffqc}. After long enough running time $\tau \gg t$ all three strategies lead to the QFI of the form
\begin{equation}
\cF_{\cE_{\tau}} = r\, \tau  \frac{\dot {s}_w^2}{\gamma_w},
\end{equation}
with different constants $r$. Optimizing the parameters $a$ and $t$ we find the values of $r$ given in Table~\ref{tab: qubit}. These values are interesting to compare with the upper-bound of Eq.~\eqref{eq: B T->0} $\cF_{\cE_{\tau}} = 16 \,\tau \frac{\dot {s}_w^2}{\gamma_w}$.  In the low temperature regime, the $\cF_{\cE_{\tau}}$ hence also scales as $T^{2\alpha}$, as seen from Eq.~\eqref{eq: B T->0}, which can be compared to similar polynomial scalings at low $T$ obtained in the literature via strong coupling~\cite{Correa_2017,Hovhannisyan_2018,Potts_2019}. These results illustrate the potential of quantum coherent control and entanglement for low-temperature thermometry. 
\begin{table}[h]
    \centering
    \begin{tabular}{cc |cc}
         strategy &  description & $r$ & $a$ \\
         \hline
         (i)& Ramsey interferometer & 1.5 & 0.5 \\
         (ii) & +entanglement with ancilla & 2.47 & 0.68 \\
         (ii) & + continuous control & $\approx$ 4.16 &  0.83 
    \end{tabular}
    \caption{The performance of the thee low-temperature Lamb shift thermometers with a qubit probe, described in the text. In all cases the QFI is given by $\cF_{\cE_{\tau}} =  r \, \tau \frac{\dot {s}_w^2}{\gamma_w}$. Initially the system is prepared in the state $\ket{\Psi_0}=\sqrt{1-a}\ket{0} + \sqrt{a} \ket{1}$ for (i), and $\ket{\Psi_0}=\sqrt{1-a}\ket{00} + \sqrt{a} \ket{11}$ for (ii) and (iii), which involve an auxiliary qubit.}
    \label{tab: qubit}
\end{table}

\subsection{Collective thermalisation}
\label{sec: Collective thermalisation}

We now consider a $N$-qubit probe, which is coupled collectively  to a  sample. The interaction Hamiltonian takes the form 
\begin{align}
  &H_I = J_x \otimes B
  \label{eq:globalCoupling}
\end{align}
with $J_x = \sum_j \sigma_x^{(j)}$, and $B$ is an unspecified operator acting on the bath.
This type of collective coupling leads to the well known phenomena of sub/superradiance \cite{dicke54},  and has received renewed interest due to the possibility of realising it in several physical platforms \cite{Kockum2019a}. It is also worth pointing out that  equilibrium thermometry with  this collective coupling  has been recently studied in \cite{Latune2020}. We now study it in the context of finite-time non-equilibrium thermometry. 

First of all, we  note that the fundamental upper bound Eq. \eqref{eq:upboundO}, obtained under Born-Markov and secular approximations, yields
\begin{align}
    \cF_{\cE_{\tau|T}} \leq \tau \left(\max_{\omega} \frac{\dot \gamma_{-\omega}^2}{\gamma_{-\omega}}\right) N^2. 
    \label{eq:upboundCol}
\end{align}
which holds independently of the specific local Hamiltonian of the probe and sample (note that the $\gamma_w$'s  depend on the specific sample into consideration). 
Hence, the coupling  \eqref{eq:globalCoupling} in principle enables a quadratic scaling with~$N$, as it may be expected from the phenomenon of superradiance.
However, saturating \eqref{eq:upboundCol} in principle requires engineering the $N$-qubit probe Hamiltonian of Eq.~\eqref{eq: optimal H} discussed in Sec. \ref{sec: optimal H}, i.e the probe needs to behave as an effective two level system (with a $2^{N-1}$ degeneracy of the upper-level). This corresponds to an $N$-body interaction between the qubits, and is a priori very hard to achieve experimentally. In what follows we consider structurally simpler probes.  

\subsubsection{Non-interacting qubits probe}

We start by considering $N$ independent qubits, i.e 
 $H=w J_z$
with  $J_z=\frac{1}{2}\sum_j \sigma^{(j)}_z$ and a tunable frequency $w$.
The  eigenstates of $H$ can be conveniently expressed in the collective spin basis. Here, it will suffice to focus on the set of symmetric states (Dicke states)\footnote{associated to the maximal value $\frac{N}{2}\left(\frac{N}{2}+1\right)$ of the total spin $\bm J^2$.}
\begin{align}
 \label{eq:superradiantstates}
     &\ket{\psi_n} =  \frac{J_+}{\sqrt{\Gamma_n}} \ket{\psi_{n-1}}, \hspace{8mm} \Gamma_n\equiv n(N+1-n)
     \nonumber\\
    & H \ket{\psi_n} = w\, (n-\frac{N}{2}) \ket{\psi_n}
 \end{align}
 with $n=0,...,N$, and where $\ket{\psi_0} = \ket{0}^{\otimes N}$ is the ground state and $J^+ = \sum_j \sigma^{(j)}_+$  (where $\sigma_+=\ketbra{1}{0}$). Constraining to the states~\eqref{eq:superradiantstates} is justified because (i) the dynamics do not mix the subspace spanned by $\{ \ket{\psi_n}\}$ with the rest of the Hilbert space, and (ii) the
   states~\eqref{eq:superradiantstates} have the fastest decay rate, and hence are the most useful for finite-time thermometry. 
 In this subspace, there are two jump operators in the master equation reading $A_w = \sum_{n=0}^{N-1} \ketbra{\psi_n}{\psi_n} J_x \ketbra{\psi_{n+1}}{\psi_{n+1}}=\sum_{n=0}^{N-1} \sqrt{\Gamma_{n+1}} \ketbra{\psi_n}{\psi_{n+1}} = J_-  $ for the transition $+w$, and $A_{-w}= A_w^{\dagger} = J_+$ for the transition $- w$ (here $J_\pm$ operators are restricted to subspace of symmetric states). Plugging these expressions in \eqref{eq: QFI bound final main}, we obtain
 \begin{align}
 \label{eq:upboundqubitsprobe}
      \frac{\cF_{\cE_{\tau|T}}}{\tau}& \leq  \left\| \sum_{n=0}^N \ketbra{\psi_n}{\psi_n}  \left[\Gamma_n \frac{\dot{\gamma}_w^2}{\gamma_w}+\Gamma_{n+1} \frac{\dot{\gamma}_{-w}^2}{\gamma_{-w}} \right] \right\|
       \nonumber\\
      &=  \frac{N}{2}\left(\frac{N}{2}+1\right) \left(\frac{\dot{\gamma}_w^2}{\gamma_w}+ \frac{\dot{\gamma}_{-w}^2}{\gamma_{-w}} \right),
 \end{align}
 where the operator norm is saturated for $n=N/2$ for which $\Gamma_{\frac{N}{2}}=\Gamma_{\frac{N}{2}+1}=\frac{N}{2}\left(\frac{N}{2}+1\right)$, for simplicity we assume an even $N$ here and in the rest of the section.
 Compared to the general upper bound \eqref{eq:upboundCol}, we also observe a quadratic scaling with $N$ but with a worse prefactor. In particular, since $0\leq  \dot{\gamma}_{w}^2/\gamma_{w} \leq \dot{\gamma}_{-w}^2/\gamma_{-w}$, we notice that for large $N$ one loses a factor laying between $1/2$ and $1/4$ as compared to the general upper bound~\eqref{eq:upboundCol}. 
 For example, for a bosonic reservoir with  rates  
 \be \label{eq: bosonic sample}
 \begin{split}
&\gamma_w = g w^\alpha N(w) \\
&\gamma_{-w} = g w^\alpha (1\pm N(w))
 \end{split}
 \ee
 and $N(w) = (e^{\beta w} - 1)^{-1}$, it is satisfied that $\dot{\gamma}_{-w}^2/\gamma_{-w}=e^{\beta w}\dot{\gamma}_{w}^2/\gamma_{w}$, so that the ratio between  \eqref{eq:upboundqubitsprobe} and  \eqref{eq:upboundCol} becomes $(1+e^{-\beta w})/4$ at leading order in $N$. It is remarkable that such a good performance can be obtained with arguably the simplest probe consisting of non-interacting qubits  with $H=w J_z $. 

\begin{figure}
\vspace{3mm}
		\centering
		\includegraphics[width=1.\linewidth]{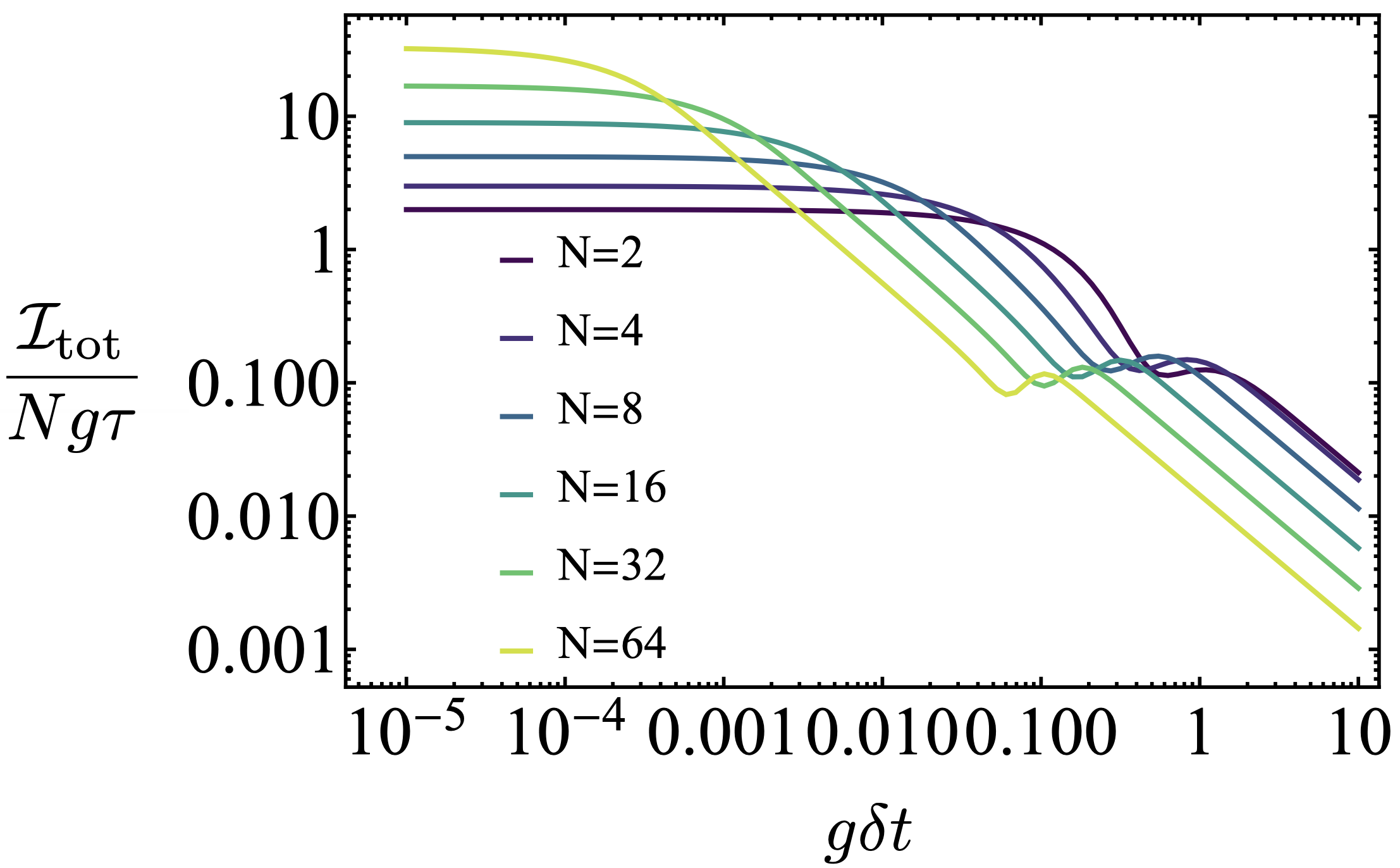}  
	    \caption{Classical Fisher information for the fast measure-and-prepare protocol where the state $\ket{\psi_{N/2}}$ evolves for a finite amount of time  $\delta t$ before performing the measurement. The superradiant advantage $\mathcal{I}\propto N^2$ requires $\delta t\propto 1/N^2$.  }
	\label{fig:fig1col}
	\end{figure}
 
 \subsubsection{The effect of measurement frequency}
 
 The bound \eqref{eq:upboundqubitsprobe} can be saturated by the fast measure-and-prepare strategy, where the probe is  continuously reinitialised it in the state~$\ket{\psi_{N/2}}$. In practice, the frequency of the measurements can be limited and it is relevant to consider suboptimal strategies where the state evolves for a finite time $\delta t$ before being measured.
 For a given $\delta t$ and total sensing time $\tau$, we then consider the fast measure-and-prepare strategy with a finite measurement frequency, described below.
 
 The probe is initialised in the state $\ket{\psi_{N/2}}$, and is left to interact with the sample for a time $\delta t$ according to the master equation
 \begin{align}
    \frac{\dd}{\dd t} \rho =& -i [H,\rho] + \gamma_w \left(J_- \rho J_+ -\frac{1}{2} \{J_+ J_-,\rho\}\right)
    \nonumber\\
    &+ \gamma_{-w} \left( J_+ \rho J_- -\frac{1}{2} \{J_- J_+,\rho\} \right).
 \end{align}
Focusing on the superradiant subspace, it is convenient to express this as a stochastic evolution of the probabilities $p_n \equiv \bra{\psi_n} \rho \ket{\psi_n}$ to occupy each state, given by
\begin{align}
    \frac{\dd}{\dd t}p_m =& - p_m \left(\gamma_w \Gamma_m +\gamma_{-w} \Gamma_{m+1}   \right)
    \nonumber\\
    & +
    \gamma_w  p_{m+1} \Gamma_{m+1} 
    + \gamma_{-w} p_{m-1}\Gamma_{m}. 
\end{align}
After a time $\delta t$, the resulting state $p_j = p_j(\delta t)$ is measured, providing the classical Fisher information $\mathcal{I}_{\delta t}=\sum_j \dot{p}_j^2/p_j$.  
 
 This process is repeated for $k$ times until $k \delta t = \tau$, yielding a total Fisher information of $\mathcal{I}_{\rm tot}=k \mathcal{I}_{\delta t} = \tau \frac{\mathcal{I}_{\delta t} }{\delta t}$. In Fig.~\ref{fig:fig1col}, we plot $\mathcal{I}_{\rm tot}/\tau ={\mathcal{I}_{\delta t} }/{\delta t}$ as a function of $\delta t$ for different $N$ and for a bosonic sample with Ohmic spectral density, $\alpha=1$ in Eq.~\eqref{eq: bosonic sample}. While  Fig.~\ref{fig:fig1col} clearly shows the quadratic advantage for $g \, \delta t \rightarrow 0$, it is also observed that this requires
 $g \, \delta t < 1/N^2$. When this condition is not met, the superradiant advantage is rapidly lost, and in fact non-superradiant states ($N=1$, prepared in the ground state) become better for  $g \delta t \geq 0.1$. 
 

 \begin{figure}
		\centering
		\includegraphics[width=1.\linewidth]{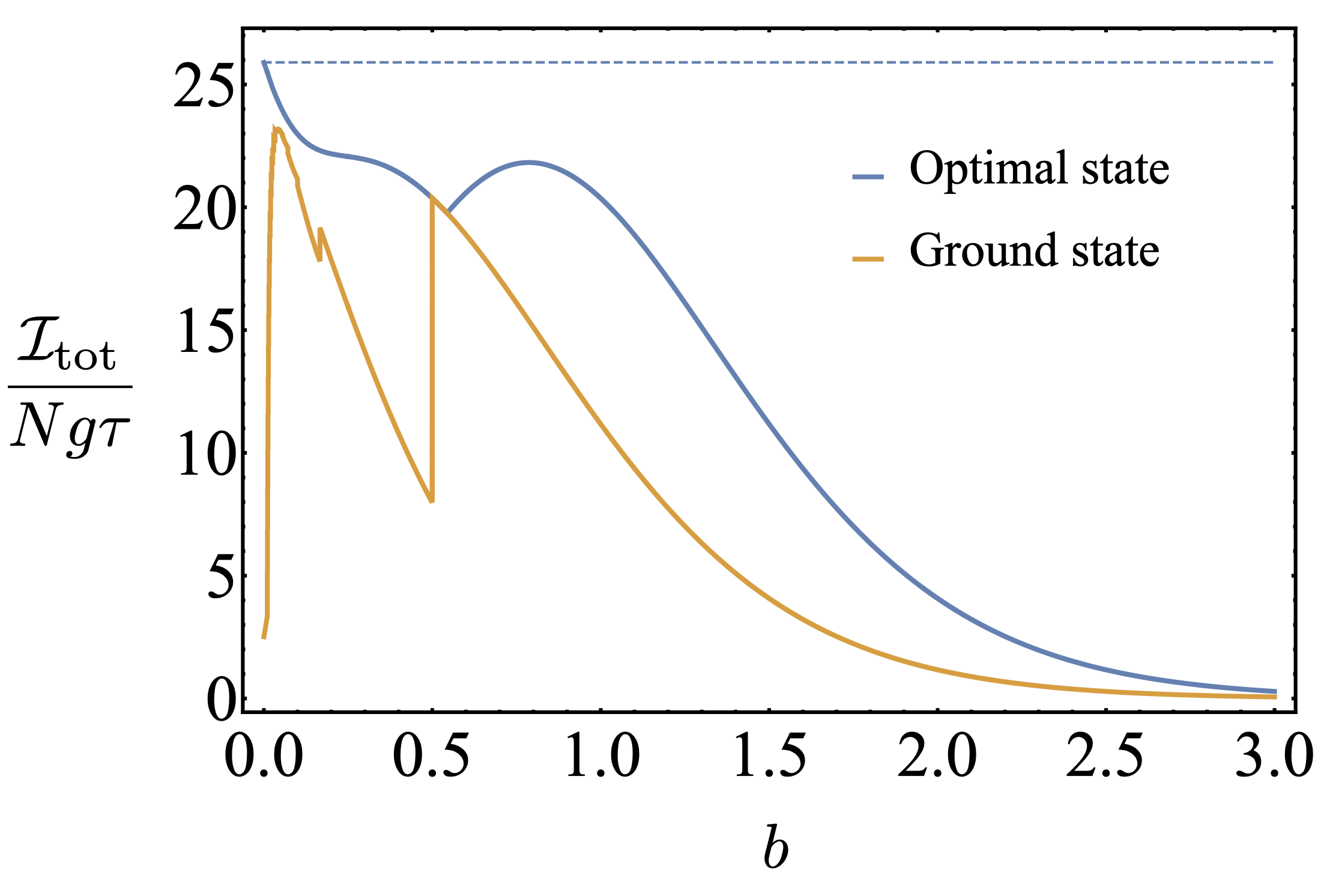} 
		\includegraphics[width=1.\linewidth]{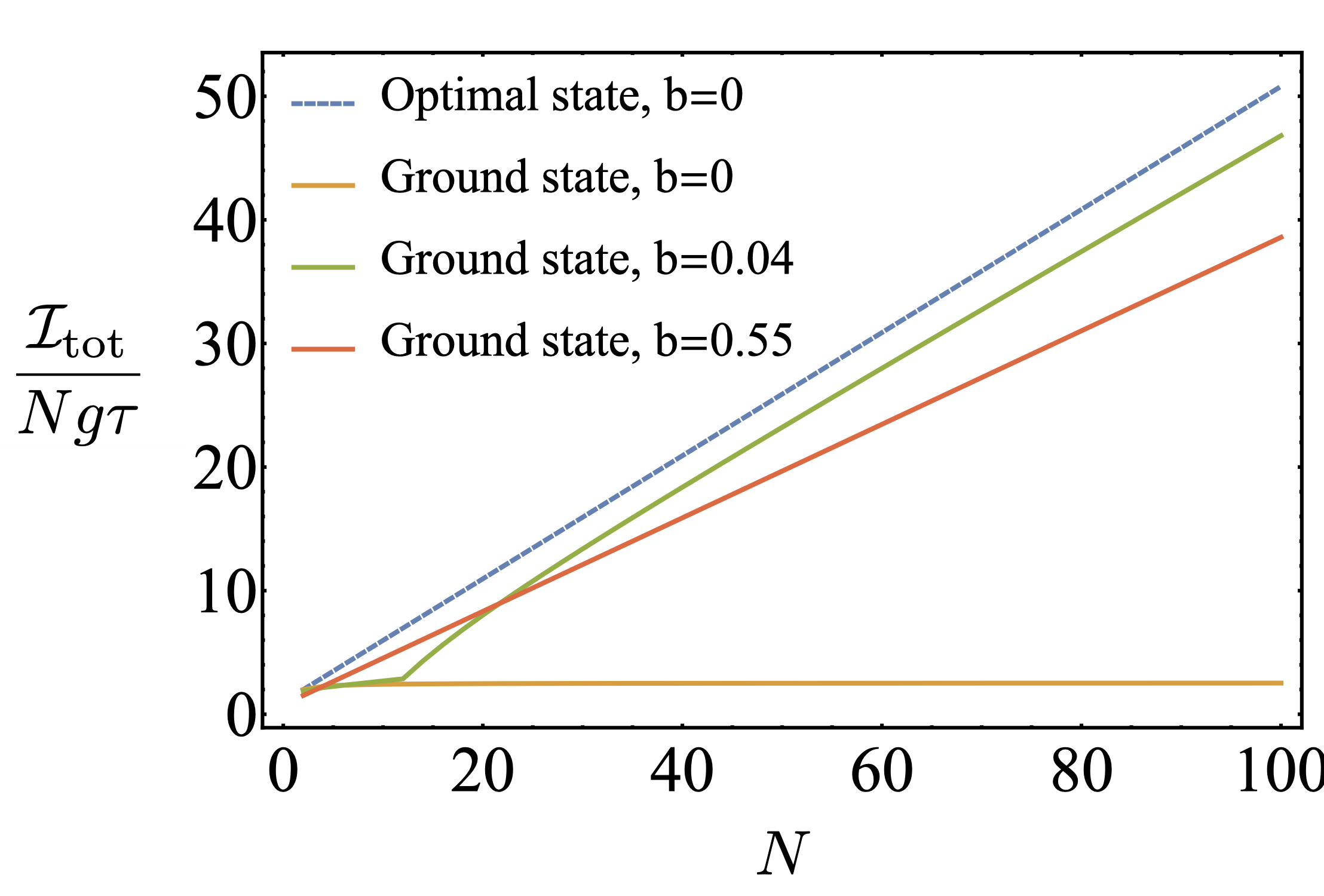} 
	    \caption{Classical Fisher information of the measure-and-prepare scheme  in the limit $\delta t \rightarrow 0$ 
	    for the generalised Hamiltonian $H(b)= \omega J_z/2 + b J_z^2$. In (a), we compare a scheme where the probe is prepared in the optimal state with one where it is prepared in the groun state. In this figure,  we vary $b$ while keeping $\omega=1$, $T=1$ and  $N=50$, and compare the results to those obtained.  In (b) we instead vary $N$ and different curves correspond to different values of $b$'s. }
	\label{fig:fig2col}
	\end{figure}

Summarising, although the simplest probe allows for a quadratic scaling of the Fisher information with $N$, this quantum advantage  faces two main challenges. (i)~The probe needs to  be repeatedly prepared in the state $\ket{\psi_{N/2}}$,
 which is highly non-trivial in practice (see e.g.~\cite{Gonzalez2017}).  (ii)~As shown in Fig.~\ref{fig:fig1col}, the time $\delta t$ after which each measurement and state preparation must be repeated decreases as $\delta t \propto 1/N^2$, which becomes increasingly difficult with increasing probe size $N$. 
  
\subsubsection{Interacting qubits probe}

We will now show that a potential route to  overcome both challenges is offered by engineering slightly more complicated probes. In particular, we now consider $N$ qubit probes subject to two-body interactions $\sigma_z^{(i)}\sigma_z^{(j)}$, the probe Hamiltonian becoming
\be \label{eq: H inter}
H= w\, J_z + b \, J_z^2,
\ee
for tunable parameters $w$ and $b$. This is arguably the simplest probe Hamiltonian after the non-interacting case (b=0) considered above, and such two-body interaction are for example implemented in spin squeezing experiments (one axis twisting~\cite{Treutlein}). The additional term $b J_z^2$ keeps the eigenvectors $\ket{\psi_n}$ unmodified, but changes the spectrum of the Hamiltonian $H\ket{\psi_n} = e_n \ket{\psi_n}$, which is now given by
 \begin{align}
     e_n = w \left(n-\frac{N}{2} \right)+ b \left(n-\frac{N}{2} \right)^2.
 \end{align}
 By deriving the energy $e_n$ with respect to $n$ to find it's local minimum, one concludes that the ground state $\ket{\psi_{n^*}}$ corresponds to 
\be
 n^* = \text{argmin}_n e_n \in \left\{\left \lfloor \frac{N}{2}-\frac{w}{2b} \right \rfloor, \left \lfloor \frac{N}{2}-\frac{w}{2b} +1\right \rfloor \right \},
 \ee
 depending on the parameter values.   It is simple to see that $n^*$ approaches $N/2$ as $b$ is increased.
 
 In particular, for $b>w$ the state $\ket{\psi_{N/2}}$ with $e_{N/2}=0$ becomes the ground state, and the energy gaps to the next levels read
 \be\begin{split}
     \Delta &= e_{N/2}- e_{N/2-1}  =-( b - w) \\
     \Delta' &= e_{N/2} - e_{N/2+1}  =-( b+ w). \\
 \end{split}
 \ee
 Our autonomous strategy of Sec.~\ref{sec: autonomous}, then gives
 \be \begin{split}
 \mathcal{I}_\text{tot} &= \tau \left ( \Gamma_{N/2} \frac{\dot \gamma_{\Delta}^2}{\gamma_{\Delta}} + \Gamma_{N/2+1} \frac{\dot \gamma_{\Delta'}^2}{\gamma_{\Delta'}} \right) \\ 
  &= \tau \, \frac{N}{2}\left(\frac{N}{2}+1\right)\left (  \frac{\dot \gamma_{\Delta}^2}{\gamma_{\Delta}} +  \frac{\dot \gamma_{\Delta'}^2}{\gamma_{\Delta'}} \right).
 \end{split}
\ee
 By choosing $w=0$ and maximizing over $b$ we get the optimal Fisher information of 
\be
 \mathcal{I}_{\text{tot}} = \frac{1}{2} \tau \left(N^2 + 2 N \right)  \max_b \frac{\dot \gamma_{-b}^2}{\gamma_{-b}},
\ee
which is only a factor $1/2$ lower as compared to the general upper-bound of Eq.~\eqref{eq:upboundCol}. Thus, engineering a quadratic Hamiltonian allows one to attain half of the optimal QFI with a simple autonomous strategy: the probe is measured and reinitialized to the ground state via its coupling to the cold bath (Fig. \ref{fig:auto}).

In practice, it might be difficult to reach a regime where~$b>w$, while getting some superradiance effect does not require going to the state $\ket{\psi_{N/2}}$.  For this reason we also numerically study the regime of lower $b$. 
In Fig. \ref{fig:fig2col} we plot $\mathcal{I}_{\rm tot}/\tau$   given a bosonic sample and setting $w=1$ and $N=50$.  Interestingly, low values of $b\approx 0.05$ provide the best performance, which is close to that obtained with  $\ket{\psi_{N/2}}$ for $b=0$ (here $w$ is not optimized but kept fixed). Furthermore, in Fig. \ref{fig:fig2col} (b), we show that this conclusion remains true as $N$ is increased.
 
In conclusion, the quadratic scaling $\mathcal{I}_{\rm tot}\propto N^2$  can in principle be realised through the autonomous scheme described in Fig. \ref{fig:auto}, where the probe is also (collectively) coupled to a $T=0$ bath. Furthermore, from the numerical example we picked one sees that this can be done even when the quadratic interaction term in the Hamiltonian of Eq.~\eqref{eq: H inter} is lower than the energy splitting of individual qubit by at least an order of magnitude $b/w < 5\%$. In principle, depending on the sample temperature $T$ and the size of the probe $N$ the required ratio $b/w$ might be much lower, but we leave this question for future work.







\section{Outlook and conclusions}

\label{sec: outlook and conclusions}

In this paper, we considered the task of estimating the temperature 
of a sample  by letting it interact with a probe for a finite duration. We derived fundamental bounds on the measurement precision  for a large class of processes where  the probe's evolution is well described by a Markovian master equation.  
This was achieved by considering the most general quantum control on the probe: arbitrary control on its state and Hamiltonian, and the possibility to realise arbitrary (adaptive, entangling) measurements. In contrast, we assumed the sample-probe interaction to be fixed,  setting a bottleneck for how fast the temperature of the sample can be imprinted on the state of the probe. In this sense, our results can be  understood as a ``causation" speed limit relating a particular sample-probe interaction with the speed at which a property of the sample causes a detectable change in the probe (quantified by the Quantum Fisher Information of the probe with respect to a particular parameter of the sample).
While deriving these results,  we realised that beyond thermometry these limits are valid in a much broader context where a physical property of a sample 
is estimated through its interaction with the probe, whose  evolution can be well described by a Markovian evolution. Our results hence provide a general framework for placing fundamental limits 
 in  finite-time sensing in open (quantum) systems.

We also discussed the attainability of 
these limits. Whenever the 
Lamb shift term in the master equation can be neglected, we showed that they can be saturated by a fast measure-and-prepare strategy, and developed an autonomous implementation of this scheme (see Fig.~\ref{fig:auto}). In principle, the presence of a temperature-dependent Lamb shift can enhance the sensitivity, which however requires moving away from the fast measurement limit to a frequency estimation regime, that we illustrated with some examples. 
This becomes particularly relevant at low temperatures, where we argued that the Lamb shift can play a dominant role for finite-time sensing (see Fig.~\ref{fig: qubit probe}). 

 The results of this paper have been obtained by assuming local control on the probe and effectively describing the thermal sample through a Markovian master equation.  This framework  is of  relevance for both  theoretical works in non-equilibrium thermometry~\cite{Mehboudi_2019rev} as well as for thermometry experiments through quantum probes~\cite{Kucsko_2013,Bouton_2020,nettersheim2022sensitivity}.   Moving beyond the Markovian approximation, a challenging future endeavour is to derive similar results in the strong coupling regime, where the probe-sample evolution is generally described  as a unitary process.  Strong coupling naturally enables a faster rate of information transfer between sample and probe, and hence appears as a promising resource for non-equilibrium thermometry, whose potential remains mostly unexplored (see Refs.~\cite{Correa_2017,Hovhannisyan_2018,DePasquale2018,Potts_2019} for works exploiting strong coupling for equilibrium thermometry). In particular, in order to generalise the results presented here, ideally one wishes  to derive fundamental bounds on the probe's QFI after a finite time for a given interaction and sample Hamiltonian. 
In general, this appears as an impossible task  when assuming full quantum control on the probe as done here. However, one may hope that some insights can be obtained 
in constrained scenarios such as in the the linear-response regime \cite{Deffner2017,Nikolai2021}. 

 When considering microscopic models for the probe-sample interaction, in this paper we focused on interactions of the form $A\otimes B$ (see Eq. \eqref{eq:Htot}). It hence remains an open and relevant question to investigate  possibilities and fundamental limits arising when considering more general probe-sample interactions of the form $\sum_j A_j \otimes B_j$. In this sense,  an exciting question that we leave for future work is whether ballistic or Heisenberg scaling is possible in thermometry, i.e., $\mathcal{F} \propto \tau^2$. 
 The fact that the temperature can also be coherently estimated through the Lamb shift as a frequency (see protocols in Sec. \ref{sec: qubit probe}) suggests that Heisenberg scaling could be possible in engineered or exotic reservoirs.


Finally, in this article we took the Quantum/Classical Fisher Information as figures of merit for thermometry. In the future, it will be interesting to consider a Bayesian framework for thermometry, and to better understand the implication of our bounds in this context.

\emph{Acknowledgments - } We thank Mohammad Mehboudi  for inspiring discussions and feedback. This research was funded by the Swiss National Science Foundation, via NCCR SwissMap and Ambizione grant PZ00P2-186067. 

 \bibliographystyle{apsrev4-1}

\clearpage
\begin{widetext}

\section{Diffusive scaling of the QFI}
\label{app: deffusive}
Consider the master equation $\frac{\dd \rho}{\dd t} \rho = \mathcal{L}(\rho)$ of Eq.~\eqref{eq: diss} with the Hamiltonian $H + H_{LS}$ and the dissipate term given by
\be
\begin{split}
    \mathcal{D}(\rho) &= \sum_\omega  \gamma_\omega \left(A_\omega\, \rho\, A^\dag_\omega - \frac{1}{2}\{A^\dag_\omega A_\omega, \rho \} \right) 
\end{split}
\ee
A Kraus representation of the infinitesimal channel $e^{\dd t \mathcal{L}}$ is given in Eq.~\eqref{eq: Kraus }
\be
\bm K = 
\begin{pmatrix}
  \id - \dd t \frac{1}{2} \sum_\omega \gamma_\omega  A_\omega^\dag A_\omega  - \ii\,  \dd t\, ( H +H_{LS})\\
  \vdots\\
 \sqrt{ \dd t \gamma_\omega } A_\omega\\
    \vdots
\end{pmatrix}.
\ee
Assuming that the jump operators are independent of the parameter, the derivatives of the Kraus operators with respect to to the parameter read
\be
\dot{\bm K} = 
\begin{pmatrix}
- \dd t\left( \frac{1}{2} \sum_\omega \dot{\gamma}_\omega  A_\omega^\dag A_\omega  + \ii \dot {H}_{LS} \right)\\
  \vdots\\
 \frac{\sqrt{ \dd t} \dot{\gamma}_\omega }{2 \sqrt{\gamma_\omega}}\, A_\omega\\
    \vdots
\end{pmatrix}.
\ee
Let us now compute $\bm K^\dag ( \dot{\bm K} +\ii h \bm K )$ up to the order $O(\dd t)$ to determine if ballistic scaling of the QFI is possible 
\be\begin{split}
   {\bm K}^\dag (\dot{\bm K}+ \ii  \, h \bm K ) &=  {\bm K}^\dag  \dot{\bm K} + \ii {\bm K}^\dag  h \bm K \\
    \\
    &=
   \begin{pmatrix}
  \id - \dd t \frac{1}{2} \sum_\omega \gamma_\omega  A_\omega^\dag A_\omega  + \ii\,  \dd t\, ( H +H_{LS})\\
  \vdots\\
 \sqrt{ \dd t \gamma_\omega } A_\omega^\dag\\
    \vdots
\end{pmatrix}^T
\begin{pmatrix}
  \dd t\left( \frac{1}{2} \sum_\omega \dot{\gamma}_\omega  A_\omega^\dag A_\omega  - \ii \dot {H}_{LS}\right)\\
    \vdots
    \\
    \frac{\sqrt{ \dd t} \dot{\gamma}_\omega }{2 \sqrt{\gamma_\omega}}\, A_\omega\\\\
    \vdots
    \end{pmatrix} +  \ii {\bm K}^\dag h \bm K \\
    &= -\dd t\left( \frac{1}{2} \sum_\omega \dot{\gamma}_\omega  A_\omega^\dag A_\omega  + \ii \dot {H}_{LS}\right) +  \dd t \frac{1}{2} \sum_\omega \dot {\gamma} _\omega A_\omega^\dag A_\omega  + \ii {\bm K}^\dag  h \bm K \\
    & = \ii \left(-\dd t   \dot {H}_{LS} + {\bm K}^\dag  h \bm K \right).
\end{split}
\ee
The last equation falls in the general context considered in \cite{Rafal2017}, from which we know that $\|-\dd t   \dot {H}_{LS} + {\bm K}^\dag  h \bm K\| = O(\dd t^{3/2})$ can be fulfilled if and only if 
\be\label{eq: span}
\dot{H}_{LS} \in \t{span} \left\{\id, \sqrt{\gamma_\omega} (A_\omega +A_\omega^\dag),\sqrt{\gamma_\omega} \ii(A_\omega -  A_\omega^\dag),  \sqrt{\gamma_\omega \gamma_{\omega'}} (A_\omega^\dag A_{\omega'}+A_{\omega'}^\dag A_{\omega}), \sqrt{\gamma_\omega \gamma_{\omega'}} \ii(A_\omega^\dag A_{\omega'}-A_{\omega'}^\dag A_{\omega}) \right\}.
\ee
One can see this explicitly by writing the Gauge matrix in the form $h =\begin{pmatrix} h_{00} & \bm h^\dag\\ \bm h & \bar h \end{pmatrix}$ with a complex vector $\bm h = (\dots h_\omega \dots)$ and an Hermitian matrix $\bar h$, so that in the leading order
\be
{\bm K}^\dag  h \bm K = h_{00} + \sum_\omega\sqrt{\dd t\gamma_\omega} (h_\omega A_\omega + h_\omega^* A_\omega^\dag) + \sum_{\omega\omega'}\dd t\sqrt{\gamma_\omega \gamma_{\omega'}} ( \bar{h}_{\omega\omega'} A_\omega^\dag A_{\omega'} +\bar{h}_{\omega\omega'}^* A_{\omega'}^\dag A_{\omega} ).
\ee\\

Let us now add the assumption the the Lamb shift Hamiltonian is block-diagonal $H_{LS} =\sum_\varepsilon \Pi_\varepsilon H_{LS} \Pi_\varepsilon$, which  brings additional structure relating the jump operators $A_\omega$ and $H_{LS}$. Recall that the jump operators are off-diagonal
\be
A_\omega = \sum_{(\varepsilon,\varepsilon')\in P_\omega} \Pi_\varepsilon A_\omega \Pi_\varepsilon',
\ee
while $H_{LS}$ (and thus $\dot{H}_{LS}$) is block-diagonal
\be
\dot{H}_{LS} = \sum_{\varepsilon} \Pi_\varepsilon \dot{H}_{LS}  \Pi_\varepsilon.
\ee
It follows that $\dot{H}_{LS}$ is orthogonal to all the jump operators (with respect to the Hilbert-Schmidt product). Furthermore, a product of two jump operators for $\omega \neq \omega'$
\be
A_\omega^\dag A_{\omega'} = \sum_{(\varepsilon,\varepsilon')\in P_\omega} \sum_{(\epsilon,\epsilon')\in P_{\omega'
}} \Pi_{\varepsilon'} A_\omega  \Pi_\varepsilon\,  \Pi_\epsilon A_{\omega'}  \Pi_{\epsilon'} = \sum_{(\varepsilon,\varepsilon')\in P_\omega} \sum_{(\epsilon,\epsilon')\in P_{\omega'
}} \Pi_{\varepsilon'} A_\omega  \Pi_\varepsilon\,  A_{\omega'}  \Pi_{\epsilon'} \delta_{\varepsilon\epsilon}
\ee
is also off diagonal $\Pi_\epsilon A_\omega^\dag A_{\omega'}\Pi_\varepsilon=0$, because each pair $(\varepsilon, \varepsilon')$ can only appear for one $\omega$. Thus $A_\omega^\dag A_{\omega'}$ for $\omega \neq \omega'$ are also orthogonal to $\dot{H}_{LS}$. Hence the only possibility for $\dot H_{LS}$ to fall in the span of the noise operators in Eq.~\eqref{eq: span} is that 
\be\label{eq: span final}
\dot{H}_{LS} \in \t{span} \left\{\id, \gamma_\omega A_\omega^\dag A_\omega \right \}
\ee
since $A_\omega^\dag A_\omega$ are the only block-diagonal contributions.\\


In the context of our microscopic model where the Lamb shift term the form
\be
\dot{H}_{LS} = \sum_\omega \dot{s}_\omega A_\omega^\dag A_\omega.
\ee
Clearly, a sufficient for the condition~\eqref{eq: span final} it is sufficient that all the rates $\gamma_\omega$ are nonzero whenever $\dot s_\omega$ is nonzero.

\section{Upper-bound on the QFI}

\label{app: QFI bound}

We now assume $\dot{H}_{LS} \in \t{span} \left\{\id, \gamma_\omega A_\omega^\dag A_\omega \right\}$ and derive an upper-bound on the QFI. Given the assumption we can write
\be
\dot{H}_{LS} = h_\id \id + \sum_\omega h_\omega  A_\omega^\dag A_\omega,
\ee
where $h_\omega =0$ for $\gamma_\omega=0$ such that the ration $\frac{h_\omega}{\gamma_\omega}$ is well defined.  Then $\| \bm K^\dag ( \dot{\bm K} +\ii h \bm K )\|=O(\dd t^{3/2})$ is fullfilled by the choice $h = \begin{pmatrix} h_{0} \dd t & \\  & \t{diag}[h_\omega/\gamma_\omega] \end{pmatrix}$. To set an upper-bound on QFI we have to compute  $\| (\dot{ {\bm K}}^\dag - \ii \bm K h)(\dot{\bm K}+ \ii  \, h \bm K )\|$. To do so, we write to the leading orders
\be\begin{split}
  (\dot{\bm K}+ \ii  \, h \bm K ) &= 
  \begin{pmatrix}
   -\dd t \left( \sum_\omega\frac{1}{2}\dot{\gamma_\omega} A^\dag_\omega A_\omega +\ii \dot{H}_{LS} \right) \\
    \vdots
    \\
    \sqrt{\dd t} \frac{\dot{\gamma}_\omega}{2 \sqrt{\gamma_\omega}} A_\omega \\
    \vdots
    \end{pmatrix}
    +
    \begin{pmatrix}
    \ii\,  h_0 \dd t\, \id \\
    \vdots
    \\
    \ii \frac{h_\omega}{\gamma_\omega} \sqrt{\dd t\gamma_\omega} A_\omega \\
    \vdots
    \end{pmatrix} 
    \\
    &=\begin{pmatrix}
   -\dd t \left( \sum_\omega (\frac{1}{2}\dot{\gamma_\omega} + \ii h_\omega \gamma_\omega) A^\dag_\omega A_\omega \right) \\
    \vdots
    \\
    \sqrt{\dd t} \frac{\dot{\gamma}_\omega + 2 \ii h_\omega}{2 \sqrt{\gamma_\omega}}  A_\omega \\
    \vdots
    \end{pmatrix} 
\end{split}.
\ee
Which makes it easy to compute
\be
(\dot{ {\bm K}}^\dag - \ii \bm K h)(\dot{\bm K}+ \ii  \, h \bm K )= \dd t \sum_\omega \left|\frac{\dot{\gamma}_\omega + 2 \ii h_\omega}{2 \sqrt{\gamma_\omega}} \right|^2 A_\omega^\dag A_\omega =  \frac{\dd t}{4} \sum_\omega \frac{\dot{\gamma}_\omega^2 + 4h^2_\omega}{\gamma_\omega}  A_\omega^\dag A_\omega .
\ee
Taking the norm gives
\be
\left\|(\dot{ {\bm K}}^\dag - \ii \bm K h)(\dot{\bm K}+ \ii  \, h \bm K ) \right\| = \frac{\dd t}{4} \left \| \sum_{\omega} \frac{\dot{\gamma}_\omega^2 + 4 h_\omega^2}{\gamma_\omega}  A_\omega^\dag A_\omega \right \|.
\ee
Using  Eq.~\eqref{eq: tau channel}, we obtain a bound on the QFI
\be
\cF_{\cE_{\tau|T}} \leq  \tau \left \| \sum_{\omega} \frac{\dot{\gamma}_\omega^2 + 4h^2_\omega}{\gamma_\omega}  A_\omega^\dag A_\omega \right \|
\ee
For $H_{LS}= \sum_\omega \dot{s}_\omega A_\omega^\dag A_\omega$ suggests a particularly simple choice of the coefficients $h_\omega = \dot{s}_\omega$, leaging to 
\be
\cF_{\cE_{\tau|T}} \leq  \tau \left \| \sum_{\omega} \frac{\dot{\gamma}_\omega^2 + 4\dot{s}_\omega}{\gamma_\omega}  A_\omega^\dag A_\omega \right \|
\ee
Finally, in the in absence of temperature dependent Lamb shift $\dot{H}_{LS}=0$, or for measurement schemes that are insensitive to it, we get the bound
\be
\cF_{\cE_{\tau|T}}\leq \tau \left \| \sum_{\omega} \frac{\dot{\gamma}_\omega^2 }{\gamma_\omega}  A_\omega^\dag A_\omega \right \|.
\ee


\section{App: Qubit probe example}
\subsection{Derivation of the optimal bound}
 To find the value of the parameter $x$ that gives the tightest bound in Eq.~\eqref{eq: bound qubit general}, we write it as
\be
\cF_{\cE_\tau|T} \leq \tau\,\min_x \text{max} \big\{f_-(x)= \frac{\dot \gamma_{w}^2+ 4 (\dot s_w +x)^2}{\gamma_{-w}},\,
f_+(x)=
\frac{\dot \gamma_w^2+ 4 (\dot s_w - x)^2}{\gamma_w} \Big\}.
\ee
The two function $f_{\pm}(x)$ are ''upwards'' parabolas centered at $x=\pm \dot s_w$, with the minimal values $ f_-(-\dot s_w)=\frac{\dot \gamma_{w}^2}{\gamma_{-w}}\geq f_+(\dot s_w) = \frac{\dot \gamma_{w}^2}{\gamma_{w}}$. To solve the minmax we distinguish  two different regimes. 

On the one hand (i), if the minimal value of $f_-(x)$, attained at $x=-\dot s_w$, is larger than $f_+(-\dot s_w)$ the best upper bound is clearly given by $f_-(-\dot s_w)$. 

On the other (ii) if $f_-(-\dot s_w)<f_+(-\dot s_w)$, the best bound is attained when the two functions are equal $f_-(x)= f_+(x)$ for $x\in [-\dot s_w, \dot s_w]$. Taking $\dot s_w \geq 0$ without loss of generality (otherwise one can flip its sign),  such an $x$ is straightforward to find 
\be\label{eq: xopt}
x_*(w)= 
-\frac{4 \dot s_w (\gamma_{w}+\gamma_{-w}) - \sqrt{64 \dot s_w^2 \gamma_{w}\gamma_{-w} - \dot \gamma_w^2 (\gamma_{w}-\gamma_{-w})^2}}{\gamma_{w}-\gamma_{-w}}.
\ee
And implies the optimal upper-bound $f_+(x_*(w)) =f_-(x_*(w))=\frac{16 \dot s_w}{\gamma_{w}-\gamma_{-w}} (-x_*(w)).$ In summary, combining (i) and (ii) we find the form of the tightest upper-bound
\be
\cF_{\cE_\tau|T}  \leq \tau 
\begin{cases}
 \frac{16 \dot s_w (-x_*(w))}{\gamma_{w}-\gamma_{-w}} & \frac{\dot\gamma_w^2}{\gamma_{-w}} < \frac{\dot\gamma_w^2 +16 \dot s^2}{\gamma_{w}}\\
\frac{\dot\gamma_w^2}{\gamma_{-w}} & \frac{\dot\gamma_w^2}{\gamma_{-w}} \geq \frac{\dot\gamma_w^2 +16 \dot s^2}{\gamma_{w}}
\end{cases}
\ee
with the expression of $x_*(w)$ given in Eq.~\eqref{eq: xopt}.

\subsection{Solving the master equation}
\label{app: ME qubit probe}

In this section we solve the master equation of the qubit probe of Sec.\ref{sec: qubit probe} in the low temperature regime $T\ll w$. We will assume a pure initial state $\ket{\Psi_0}$. The master equation takes the form
\be\label{eq: qubit ME app}
\frac{\dd}{\dd t} \rho = -\ii \big[\frac{1}{2} \tilde w \, \sigma_z, \rho\big] +\gamma_{w}\left( \underbrace{ \sigma_- \rho\, \sigma_+ -\frac{1}{2}\big\{\prjct{1},\rho\big\}}_{=\mathcal{D}(\rho)} \right).
\ee
with $\tilde w = w + 2 \Delta_T + \Delta$. 
One notes that in this equation the unitary dynamics commute with the dissipation $[\sigma_z,\mathcal{D}(\rho)]=\mathcal{D}\left([\sigma_z,\rho]\right)$. Hence, we can decompose the evolution as ''pure'' dissipation $\frac{\dd}{\dd t} \rho = \gamma_{w}\mathcal{D}(\rho)$ acting on the state for time $t$ followed by unitary evolution $\frac{\dd}{\dd t} \rho = - \frac{\ii}{2} \tilde w [\sigma_z,\rho]$ of the same duration. The unitary dynamics results in the unitary operator
\be
U_t = \exp(  -\frac{\ii}{2} t\,  \tilde w\,  \sigma_z).
\ee
The stochastic dynamics projects the system on the ground state $\prjct{0}\propto \sigma_- \rho(t) \sigma_+ $ with the rate $\tr \prjct{1} \rho(t)$. At each time it gives result to two possible branches, either the jump $\sigma_-$ has already occurred and the system remains in $\rho_\text{j}(t)=p_\text{j}(t) \prjct{0}$, or it has not yet occurred $\rho_\text{nj}(t)$. In the no-jump branch the dynamics is given by $\frac{\dd}{\dd t} \rho_\text{nj}(t) = -\frac{\gamma_w}{2}\{\prjct{1},\rho_\text{nj}(t) \}$, and is straightforward to solve
\be
\rho_\text{nj}(t)  =  \prjct{\Psi_\text{nj}(t)} \qquad \ket{\Psi_\text{nj}(t)} = e^{-\frac{t}{2}\gamma_\omega \prjct{1}} \ket{\Psi_0}
\ee
The probability that a jump has occurred before $t$ is thus 
\be
p_\t{j}(t) = 1 - \tr \rho_\text{nj}(t) = 1- \tr e^{-\gamma_w t \prjct{1}} \rho_0.
\ee
Plugging in the coherent evolution given by $U_t$, the state at time $t$ is
\be\label{eq: probe qubit solution app}
\rho_t = U_t \rho_\mathcal{D}(t) U_t^\dag \qquad \rho_\mathcal{D}(t) = p_\text{j}(t) \prjct{0} +\rho_\text{nj}(t) 
\ee 
We compute the QFI of the state as $\cF_{\rho_t} = \tr \dot \rho_t L$ where $L$ is the symmetric logarithmic derivative $\frac{1}{2}\{\rho_t,L\}=\dot \rho_t$. In the frame rotating with $U_t$ one has
\be \label{eq: SLD guys qubit}
\rho_t = \rho_\mathcal{D}(t) \qquad \dot \rho_t = -\frac{\ii}{2} t \dot{\tilde w} [\sigma_z,\rho_\mathcal{D}(t) ],
\ee
where $\dot{\tilde w} = 2 \dot \Delta_T = 2\dot s_w$.

\subsection{Ramsey scheme}

Let us now consider a measurement scheme where the probe is prepared is subject to free evolution in Eq.~\eqref{eq: qubit ME app} for time $t$, after which it is mearused and reinitialised. Without loss of generality we assume that the probe is prepare in a state in th x-z plane of the Bloch sphere $\ket{\Psi_0}= \sqrt{1-a}\ket{0} +\sqrt{a}\ket{1}$, such that
\be
\rho_0 = \begin{pmatrix}
1-a & \sqrt{a-a^2} \\
\sqrt{a-a^2} & a
\end{pmatrix} = (\sqrt{1-a} \ket{0} + \sqrt{a}\ket{1})(\sqrt{1-a} \bra{0} + \sqrt{a}\bra 1)
\ee
with $a\in[0,1]$. From Eqs.~(\ref{eq: probe qubit solution app}-\ref{eq: SLD guys qubit}) we get 
\be\begin{split}
\rho_t =  \begin{pmatrix}
1-e^{-t \gamma_w} a & e^{-\frac{\gamma_w}{2} t} \sqrt{a-a^2} \\
e^{-\frac{\gamma_w}{2} t} \sqrt{a-a^2} & e^{-{\gamma_w} t}a 
\end{pmatrix} \qquad
\dot \rho_t = 2\dot s_w t e^{-\frac{\gamma_w}{2} t} \sqrt{a-a^2}  \sigma_y.
    \end{split}
\ee
With these expressions one verifies that the SLD is given by  $L = \dot s_w t e^{-\frac{\gamma_w}{2} t} \sqrt{a-a^2} \sigma_x$ and the QFI
$\cF_{\rho_t} = \tr \dot \rho_t L = 16 {\dot s}_w^2 t^2 e^{-\gamma_w t} (a-a^2).$
The QFI is maximized to
\be
\cF_{\rho_t} = 4 {\dot s}_w^2 t^2 e^{-\gamma_w t}
\ee
for $\rho_0 =\prjct{+}$, i.e. $a=\frac{1}{2}$.  
It remains to determint the optimal duration between the preparation and the measurement $t_\t{opt}$, i.e.  which maximizes the QFI rate 
\be
t_\t{opt} = \t{argmax}_t\,  \frac{\cF_{\rho_t} }{t} = \frac{1}{\gamma_w}.
\ee
The best strategy is then to measure the state of the probe after $t_\t{opt}$, and reinitialize it in the state $\ket{+}$ after the measurement. For a long enough running time given the QFI becomes
\be
\cF_{\cE_{\tau}} = \tau\,  \frac{\cF_{\rho_{t_\t{opt}}} }{t_\t{opt}} = \tau \, \frac{4}{e} \frac{\dot {s}_w^2}{\gamma_w} \approx 1.5\, \tau   \frac{\dot {s}_w^2}{\gamma_w} 
\ee

In comparison, the upper-bound of Eq.~\eqref{eq: B T->0} reads $\cF_{\cE_{\tau}} \leq 16 \tau \ \frac{\dot {s}_w^2}{\gamma_w} $ in the limit $\dot s_w \gg \dot \gamma_w$, and is larger by an order of magnitude.\\

\subsection{Error detection scheme}

Next, let us consider the case where the evolution is subject to quantum control. In particular, we introduce an auxiliary and prepare the initial state
\be
\ket{\Psi_0} = \sqrt{1-a}\ket{00} + \sqrt{a} \ket{11}.
\ee
First, consider the case where no control operations are performed on the state between the preparation and the measurement. The Lindbladian in Eq.~\eqref{eq: qubit ME app} acts trivially on the auxilliary qubit, and the evolution of the probe and ancilla system is thus almost identical with the case described above. The only difference is that the jump operator $\sigma_-$  (acting on the probe only) leaves the system in the state
\be
(\sigma_-\otimes \id) \rho(t) (\sigma_+\otimes \id) \propto \prjct{0}\otimes \prjct{1},
\ee
which has a different \emph{parity} as compared to the initial state. While in the no-jump branch the parity of the initial state is preserved, i.e. in the jump branch the state evolves to
\be
\ket{\Psi_\text{nj}'(t)} = \sqrt{1-a}\ket{00} + \sqrt{a} e^{-\frac{\gamma_w}{2} t } \ket{11},
\ee
And the final state is now a direct sum
\be
\rho_\mathcal{D}(t) = \prjct{\Psi_\text{nj}'(t)} \oplus p_\t{j}(t) \prjct{01}
\ee
rather then a mixture. Futhremore, note that the coherenet evolution $U_t = \exp\left(-\frac{\ii}{2}t\, \tilde w \sigma_z\otimes\id\right)$, preserves the parity. Hence, by measuring the parity  $\{\Pi_\checkmark, \Pi_{\cross}\}$, with 
\be\begin{split}
\Pi_\checkmark &= \prjct{00}+\prjct{11} \\
\Pi_{\cross} &= \prjct{01}+\prjct{10},
  \end{split}
\ee
of the two-qubit state at any time,
one can learn if a jump has occurred without disturbing the coherent part of the evolution. From 
\be\begin{split}
\rho_t = \prjct{\Psi_\text{nj}'(t)} \oplus p_\t{j}(t) \prjct{01} \qquad
\dot \rho_t = 2\dot s_w t e^{-\frac{\gamma_w}{2} t} \sqrt{a-a^2} \, \ii (\ketbra{11}{00}-\ketbra{00}{11})
    \end{split}
\ee
we compute
\be\label{eq: (ii) QFI}
\cF_{\rho_t} = 16 \frac{(1-a)a t^2 {\dot s}_w^2}{(1-a)e^{t \gamma_w} +a}.
\ee
To find the optimal initial state $a$ and sensing time $t$, we again optimize the QFI rate 
\be
\cF_{\cE_\tau} = \tau \max_{a,t} \frac{\cF_{\rho_t}}{t}  \approx 2.47\, \tau   \frac{\dot {s}_w^2}{\gamma_w},
\ee
and obtained for an initial state with $a\approx 0.68$. We see that in improves  over the previous strategy.

\subsection{Fast error detection scheme}

As the parity measurement commutes with the evolution, a more involved control strategy is to monitor the occurrence of jumps all the time. In this case when an error occurs, the state of the system is projected to $\ket{0,1}$ and carries no information on the temperature (recall that we neglect the dependence of the jump rate on the temperature $\dot \gamma_w\approx 0$ which is exponentially suppressed). If a jump is detected one reinitialises the state of the probe and auxiliary qubits immediately. This allows one to save some ''sensing'' time.

The probability that no jumps occurs between $t=0$, when the state $\ket{\Psi_0} = \sqrt{1-a}\ket{00} + \sqrt{a} \ket{11}$ is prepare, and $t$ reads
\be
p_\t{nj}(t) = \tr \prjct{\Psi_\t{nj} (t)} = 1-a + a \, e^{-\gamma_w t}.
\ee
While the probability that the first jump happens during the infinitesimal time interval $[t,t+\dd t]$ is
\be
r_\t{j}(t) = - \frac{\dd }{\dd t} p_\t{nj}(t) \dd t  = a \gamma_w e^{-\gamma_w t}.
\ee
Hence, if the strategy is to wait for a time $T$ before measuring the system conditional on no-errors, the average duration of a single trial is 
\be
\mean{t}_T = p_\text{nj}(T) T + \int_{0}^T \dd t\,  r_\t{j}(t) t = \frac{a \left(1 -e^{-\gamma_w T} \right) + (1-a) T \gamma _w}{\gamma _w}.
\ee
It is the average duration of the between the preparation and the first reinitializtion. In addition we can compute the average QFI of a single trial. The only nonzero contribution to QFI, comes from the branch where no error occurs and we have already computed it in the previous section
\be
\mean{\cF}_T = 16 \frac{(1-a)a T^2 {\dot  s}_w^2}{(1-a)e^{T \gamma_w} -a}.
\ee
For long sensing times we can estimate the QFI rate by the ratio between the average QFI and the average duration of a trial
\be
\cF_{\cE_\tau } \approx \tau \max_{a,T} \frac{\mean{\cF}_T}{\mean{t}_T} \approx 4.16\,   \tau   \frac{\dot {s}_w^2}{\gamma_w},
\ee
obtained for $a\approx0.83$.
\end{widetext}
\end{document}